\documentclass[floatfix, twocolumn, prb]{revtex4-2}

\usepackage{multirow}
\usepackage{epsfig}
\usepackage{graphicx}
\usepackage{amsmath}
\usepackage{amssymb}
\usepackage{bm}
\usepackage{float}
\usepackage{dcolumn}
\usepackage{braket}
\usepackage{bbold}
\usepackage{ulem}
\usepackage{soul}
\usepackage{rotating}
\usepackage{tikz,pgfplots,gensymb}
\usepackage[hidelinks]{hyperref}
\usepackage{xcolor}
\usepackage{pifont}
\usepackage{array}
\usepackage{natbib}
\usepackage{longtable}
\usepackage{color, colortbl}

\definecolor{Gray}{gray}{0.9}
\definecolor{LightCyan}{rgb}{0.88,1,1}
\definecolor{Blue}{rgb}{0,0,1}

\pgfplotsset{compat=1.8}
\usetikzlibrary{pgfplots.groupplots}

\hypersetup{
    colorlinks,
    linkcolor={red!50!black},
    citecolor={blue!50!black},
    urlcolor={blue!80!black}
}

\begin{document}

\title{Enhanced anomalous Nernst effects in ferromagnetic materials driven by Weyl nodes}

\author{Ilias Samathrakis}
\email[corresp.\ author: ]{iliassam@tmm.tu-darmstadt.de}
\author{Teng Long}
\author{Zeying Zhang}
\author{Harish K. Singh}
\author{Hongbin Zhang}

\affiliation{Institute of Materials Science, TU Darmstadt, 64287 Darmstadt, Germany}

\begin{abstract}
Based on high-throughput first-principles calculations, we evaluated the anomalous Hall and anomalous Nernst conductivities of 266 transition-metal-based ferromagnetic compounds. Detailed analysis based on the symmetries and Berry curvatures reveals that the origin of singular-like behaviour of anomalous Hall/Nernst conductivities can be mostly attributed to the appearance of Weyl nodes or nodal lines located in the proximity of the Fermi energy, which can be further tailored by external stimuli such as biaxial strains and magnetic fields. Moreover, such calculations are enabled by the automated construction of Wannier functions with a success rate of 92\%, which paves the way to perform accurate high-throughput evaluation of the physical properties such as the transport properties using the Wannier interpolation.

\end{abstract}

\maketitle

\section{\label{INTRO}Introduction}

In the last decades, materials of the nontrivial topological nature have attracted quite intensive attention of the community, 
leading to many interesting physical properties such as ultra high mobility~\cite{liang2015ultrahigh}, protected surface states~\cite{xu2015discovery}, anomalous magnetoresistance~\cite{huang2015observation,liang2015ultrahigh} and exotic optical properties~\cite{wu2017giant}, promising for future applications. One particularly interesting class of materials are those with long range magnetic orderings at finite temperatures, including topological insulators, Dirac (Weyl) semimetals (DSM,WSM) and nodal line semimetals. These materials are characterized by gapless surface Dirac fermions with a spin-momentum-locked energy dispersion, a spin-(non-)degenerate band touching point and lines respectively. \cite{zhangHighthroughputDesignMagnetic2020,satoTopologicalSuperconductorsReview2017} These special energy dispersions are responsible the generation of interesting physical phenomena. For instance, for ferromagnetic materials, the anomalous Hall effect (AHE), which describes the generation of a transversal current perpendicular to the electric current in the absence of an external magnetic field~\cite{vzutic2004spintronics,nagaosa2010anomalous}, can be induced. The same can be achieved using the thermal gradient as the external stimulus, resulting in the anomalous Nernst effect (ANE)~\cite{behnia2016nernst,lee2004anomalous}. Both AHE and ANE are caused by the nonzero Berry curvature of the Fermi sea states,~\cite{nagaosa2010anomalous} which acts as a fictitious magnetic field~\cite{xiao2010berry} giving rise to these effects that lead to promising applications such as data storage and data transfer.~\cite{hasegawa2015material,huang2011intrinsic} Note that like the other linear response properties, the anomalous Hall conductivities (AHC) can also be equally evaluated by integrating the Fermi surfaces~\cite{wang2007fermi}, arising one interesting aspect whether AHC can be tuned by external stimuli.

Recently, motivated by the emergent antiferromagnetic spintronics~\cite{baltzAntiferromagneticSpintronics2018}, it is demonstrated that significant AHCs and ANCs can be obtained in magnetic materials with noncollinear antiferromagnetic ordering~\cite{Suzuki:2017, seemann2015symmetry}. The most representative materials include the Mn$_3$X with X=Ge,Ga,Sn,Rh,In,Pt family \cite{Chen:2014,YZhang:2018,Nayak:2016,ganguly2011augmented,yang2017topological,Kubler:2014,li2017anomalous,Guo:2017,Kiyohara:2016,ikhlas2017large,Zelezny:2017}, the Mn$_3$XN with X=Ga,Zn,Ag,Ni family~\cite{Zhou:2020,Gurung:2019,Huyen:2019}, Mn$_5$Sn$_3$~\cite{Surgers:2017,Surgers:2018} and Pr$_{2}$Ir$_{2}$O$_{7}$~\cite{Machida:2007}. Besides AHC, the triangular noncollinear antiferromagnetic structure of the Mn$_3$X family is responsible for the generation of a spin current in the absence of external field, dubbed Spin Hall effect (SHE)~\cite{zhangStrongAnisotropicAnomalous2017} and the Magneto-optical Kerr effect (MOKE)~\cite{higoLargeMagnetoopticalKerr2018} which is the rotation of the plane of polarization of a light beam reflected from the surface in the presence of a magnetic field. Such peculiar properties can be traced back to the existence of Weyl nodes in the electronic structure.~\cite{yang2017topological,vsmejkal2018topological} WSM are topological materials where the electronic structure is dominated by non-accidental linear touching and historically Y$_2$Ir$_2$O$_7$ was the first material to be realized as a WSM~\cite{wanTopologicalSemimetalFermiarc2011}. Recently, Co$_3$Sn$_2$S$_2$ and several magnetic Weyl Heusler compounds were classified as WSM which also exhibit touching points with linear dispersion close to the Fermi level~\cite{armitage2018weyl,yan2017topological}, giving rise to large AHC. Therefore, there is a strong impetus to search for more magnetic WSM and to investigate the resulting physical properties.

In this work, after collecting the known ferromagnetic materials, we performed high-throughput (HTP) calculations on AHC and ANC of  266 existing transition-metal-based ferromagnetic compounds, where the magnetization directions and biaxial strains are considered as perturbations to tune such effects. It is observed that the appearance of linearly degenerate states such as Weyl nodes and nodal lines will lead to singular-like behavior of AHC and hence enhanced ANC. Nevertheless, there exist also materials with significant ANC but without hot-spot contributions to AHC. Therefore, it is suspected that further HTP calculations are required in order to identify promising candidates with enhanced ANC for transversal thermoelectric applications, where automated construction of Wannier functions can be valuable.

\section{Numerical details}

Out of 5487 experimentally known ferromagnets, reported in AtomWork Adv database~\cite{atomwork}, the crystal structure of 3956 was found and collected using ICSD,~\cite{icsd} Materials Project,~\cite{materialsproject} and AtomWork databases. Compounds being either non-stoichiometric or containing elements with partial occupation were excluded, leading to 1827 compounds. Additionally, rare earths, oxides and compounds including more than 30 atoms in their unit cell were screened out as well since they increase the complexity and the time needed for the calculation. These constraints further reduced the number of available ferromagnetic compounds for calculations to 335.

The transport properties of the ferromagnetic compounds are computed automatically using an in-house-developed scheme, written in Python, linking VASP, Wannier90 and Wanniertools software. The first step involves the self consistent first-principles calculation of each compound using the projected augmented wave method (PAW), implemented in VASP.~\cite{Kresse:1993} The exchange correlation functional is approximated in the general gradient approximation (GGA), as parameterized by Perdew-Burke-Ernzerhof~\cite{Perdew:1996} and the spin orbit coupling is included in all calculations. A $\Gamma$-centered kmesh of density 50, in respect to the lattice parameters, as well as an energy cutoff of 500$eV$ are selected. Subsequently, the Bloch wave functions are projected onto maximally localised wannier functions (MLWF), using Wannier90, following Ref.~\onlinecite{Mostofi:2008}. The number of MLWF as well as the disentanglement and frozen windows of each compound are automatically computed following Ref.~\onlinecite{Zeying:2018}. The AHC is then evaluated by integrating the Berry curvature, according to the formula:
\begin{align}
& \sigma_{\alpha \beta} = - \frac{e^2}{\hbar} \int \frac{d\mathbf{k}}{\left( 2\pi \right)^3} \sum_{n} f \big[ \epsilon \left( \mathbf{k} \right) -\mu \big] \Omega_{n,\alpha \beta} \left( \mathbf{k} \right) \label{ahc} \\
& \Omega_{n,\alpha \beta}\left( \mathbf{k} \right) = -2Im \sum_{m \neq n} \frac{\braket{\mathbf{k}n|v_{\alpha}|\mathbf{k}m}\braket{\mathbf{k}m|v_{\beta}|\mathbf{k}n}}{\big[ \epsilon_m\left( \mathbf{k} \right) - \epsilon_n\left( \mathbf{k} \right) \big]^2} \label{bc},
\end{align}
where $f\big[ \epsilon \left( \mathbf{k} \right) - \mu \big]$ denotes the Fermi distribution function, $\mu$ the Fermi energy, n (m) the occupied (empty) Bloch band, $\epsilon_n \left( \mathbf{k} \right) \left( \epsilon_m \left( \mathbf{k} \right) \right)$ their energy eigenvalues and $v_{\alpha}$ $\left( v_{\beta} \right)$ the velocity operator. The integration is performed on 400 dense kmesh in respect to the lattice parameters, using Wanniertools.~\cite{wanniertools:2018} The ANC is defined according to the formula:
\begin{equation}
a_{\alpha \beta} = -\frac{1}{e} \int d\epsilon \frac{\partial f}{\partial \mu} \sigma_{\alpha \beta} \left(\epsilon\right) \frac{\epsilon - \mu}{T},
\end{equation}
where $\epsilon$ is the energy point of the energy grid, e the electron's charge, T the temperature and $\mu$ the Fermi level. Regarding the ANC, an in house developed Python script in an energy grid of 1000 points at a range $\big[-0.5,0.5\big]$ eV in respect to the Fermi level is used.

\section{\label{RES}Results and Discussion}

As stated above, starting from 335 transition-metal-based ferromagnetic materials, we managed to converge the DFT calculations for 289 compounds for both magnetization directions, {\it i.e.}, the [001] and [100]. MLWFs can be successfully constructed for 266 of them, giving rise to a success rate of 92\%. In comparison, a success rate of 93\% and 97\% is achieved in Ref.~\onlinecite{garrity2021database} and Ref.~\onlinecite{vitale2020automated} respectively. Our results provide an alternative to the automatic construction of  Wannier functions and enable us to perform further HTP calculations to evaluate the desired physical properties, in particular making use of the Wannier interpolation technique~\cite{marzariMaximallyLocalizedWannier2012}.

\begin{figure*}
\includegraphics[width=0.8\textwidth]{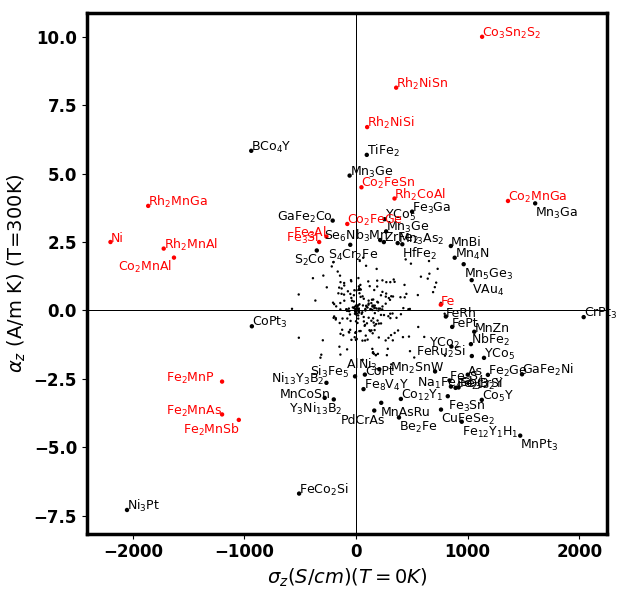}
\caption{$\sigma_z$ (at T=0K) and $\alpha_z$ (at T=300K) for selected ferromagnets with the magnetization direction parallel to the [001] axis. Our results are indicated with black and compared to the ones available in literature in red. 
}
\label{fig1}
\end{figure*}

Figure.~\ref{fig1} shows the z-component ($\sigma_z=\sigma_{xy}$) of AHC and ANC for 266 FM intermetallic compounds with the magnetization aligned along the [001] direction. Obviously, the magnitude of AHC ranges between -2051$S/cm$ for Ni$_3$Pt and 2040$S/cm$ for CrPt$_3$, where there are 11 compounds with the absolute value of AHC exceeding 1000$S/cm$. The largest magnitude of AHC is larger than the already reported -1862$S/cm$ for Rh$_2$MnGa and -1723$S/cm$ for Rh$_2$MnAl,~\cite{noky2020giant} arising the question whether there is an upper limit and opens up the possibility to engineer materials with more significant magnitudes. Concomitant with AHC, the z-component ($\alpha_z=\alpha_{xy}$) of ANC varies between -7.29$A/(m \cdot K)$ for Ni$_3$Pt and 5.83$A/(m \cdot K)$ for BCo$_4$Y, and exceeds 3$A/(m \cdot K)$ in 16 compounds. 

In general, our results match reasonably well with the existing literature for the magnetization direction being parallel to z-axis. More specifically the reported value of 1130$S/cm$ for Co$_3$Sn$_2$S$_2$~\cite{liu2018giant} is comparable with the 988$S/cm$ obtained from our calculations. Moreover, good agreement is achieved in Heusler compounds where our values of 140$S/cm$ and 200$S/cm$ for Co$_2$VGa and Co$_2$MnSn agree with the reported cases of 137$S/cm$ in Ref.~\onlinecite{manna2018colossal} and 118$S/cm$ in Ref.~\onlinecite{kubler2012berry} respectively. Comparable results are also observed in the XPt$_3$ family with $X=\left(Cr,Mn\right)$ where the reported cases of 2040$S/cm$ and 1400$S/cm$~\cite{markou2020hard} are reproduced by our values of 2000$S/cm$ and 1471$S/cm$ respectively. Regarding ANC, the calculated value of 4.58$A/(m \cdot K)$ for MnPt$_3$ is consistent with the reported value of 4$A/(m \cdot K)$ in Ref.~\cite{markou2020hard}. It is important to mention that the ANC is extremely sensitive to the density of points used for the calculation as well as the temperature that might explain several differences between the reported and the calculated values. 

The shape of the AHC and ANC tensors is the same and can be determined by the magnetic space group of the compound. For the AHC tensor given by Eq.~\ref{ahc}, it primarily depends on the Berry curvature which behaves as a pseudovector under symmetry operations, yielding:
\begin{equation}
s\mathbf{v\left(r\right)}=\pm det \left( \mathbf{D}\left( R \right) \right) \mathbf{D}\left( R \right) \mathbf{v}\left(s^{-1}\mathbf{r}\right)
\label{symtrans}
\end{equation}
\noindent where $\mathbf{v}\left(r\right)$ denotes the pseudovector Berry curvature, $\mathbf{D}\left( R \right)$ the three-dimensional representation of a symmetry operation without the translation part and $s$ an arbitrary symmetry operation. 
Taking SiMnY as an example, we observe that its crystal structure belongs to the space group $P4/nmm$ (129) and the magnetization direction along the [100]-direction renders it to the magnetic space group is $Pmm'n'$ (BNS: 59.410). 
Correspondingly, the Berry curvature (in the Cartesian basis) behaves as an odd pseudovector under the application of $m_{100}$ and obeys:
\begin{align}
&\Omega_x\left(-k_x,k_y,k_z\right)=\Omega_x\left(k_x,k_y,k_z\right) \notag \\
&\Omega_y\left(-k_x,k_y,k_z\right)=-\Omega_y\left(k_x,k_y,k_z\right) \notag \\
&\Omega_z\left(-k_x,k_y,k_z\right)=-\Omega_z\left(k_x,k_y,k_z\right) \notag.
\end{align}
Thus, the summation over the whole Brillouin zone forces $\sigma_y=\sigma_{xz}$ and $\sigma_z=\sigma_{xy}$ to vanish. 
On the other hand, the magnetic space group of SiMnY changes to $P4/nm'm'$ (BNS: 129.417) for the magnetization direction parallel to [001] axis. In this case, the Berry curvature behaves as an odd pseudovector under the application of $2_{100}$ and obeys
\begin{align}
&\Omega_x\left(-k_x,-k_y,k_z\right)=-\Omega_x\left(k_x,k_y,k_z\right) \notag \\
&\Omega_y\left(-k_x,-k_y,k_z\right)=-\Omega_y\left(k_x,k_y,k_z\right) \notag \\
&\Omega_z\left(-k_x,-k_y,k_z\right)=\Omega_z\left(k_x,k_y,k_z\right) \notag.
\end{align}
It is obvious that the summation over the whole Brillouin zone forces $\sigma_x=\sigma_{yz}$ and $\sigma_y=\sigma_{xz}$ to vanish, leading to the conclusion that for the high-symmetric magnetization directions, the direction of the AHC tensor is always aligned with the magnetization direction. It is further noted that the absence of symmetries dictating specific components of the Berry curvature to be zero does not necessarily guarantee finite AHC of the respective component, since AHC depends on the distribution of the Berry curvature in the whole BZ. For instance, it is calculated negligible AHC value for all components in Co$_2$NbAl ($Im'm'm$, BNS:71.536) despite the absence of symmetry forcing $\sigma_x=0$. 

For magnetic materials without time-reversal symmetry, the actual symmetry and the actual electronic structure depend on the magnetization directions after considering SOC, leading to in general anisotropic responses. In order to quantify the changes in AHC and ANC, we performed calculations on two different magnetization directions, {\it i.e.}, [100] and [001]. The resulting anisotropy can be expressed as the ratio of the so obtained values for the two magnetization directions, {\it i.e.}, $\sigma_{x,M\parallel[100]}/\sigma_{z,M\parallel[001]}$ or $\alpha_{x,M\parallel[100]}/\alpha_{z,M\parallel[001]}$. It is clear that ``large'' values signify large changes in favor of the [100] direction, as it happens for AHC and ANC in Rh$_2$MnSb and Fe$_3$Se$_4$, where anisotropy of 53.32 and -33.39 is calculated respectively. On the other hand, small values (excluding the cases where both AHC values are lower than 10$S/cm$ and both ANC values lower than 0.05$A/(m \cdot K)$ and therefore considered negligible) are in favor of the [001] magnetization direction, such as -0.01 and 0.004, that are present in PFe and Fe$_2$Ge for AHC and ANC respectively. The full list of the values obtained for AHC, ANC and anisotropy for all ferromagnetic compounds are available in Table S1 of the supplementary\cite{supplementary}. Compounds with AHC larger than or equal to 1000$S/cm$ (irrespective of the direction), ANC larger than or equal to 3$A/(m \cdot K)$ and anisotropy larger than 4 or smaller than 0.25 are classified as extreme cases and they are highlighted.

\begin{figure*}
\includegraphics[width=0.8\textwidth]{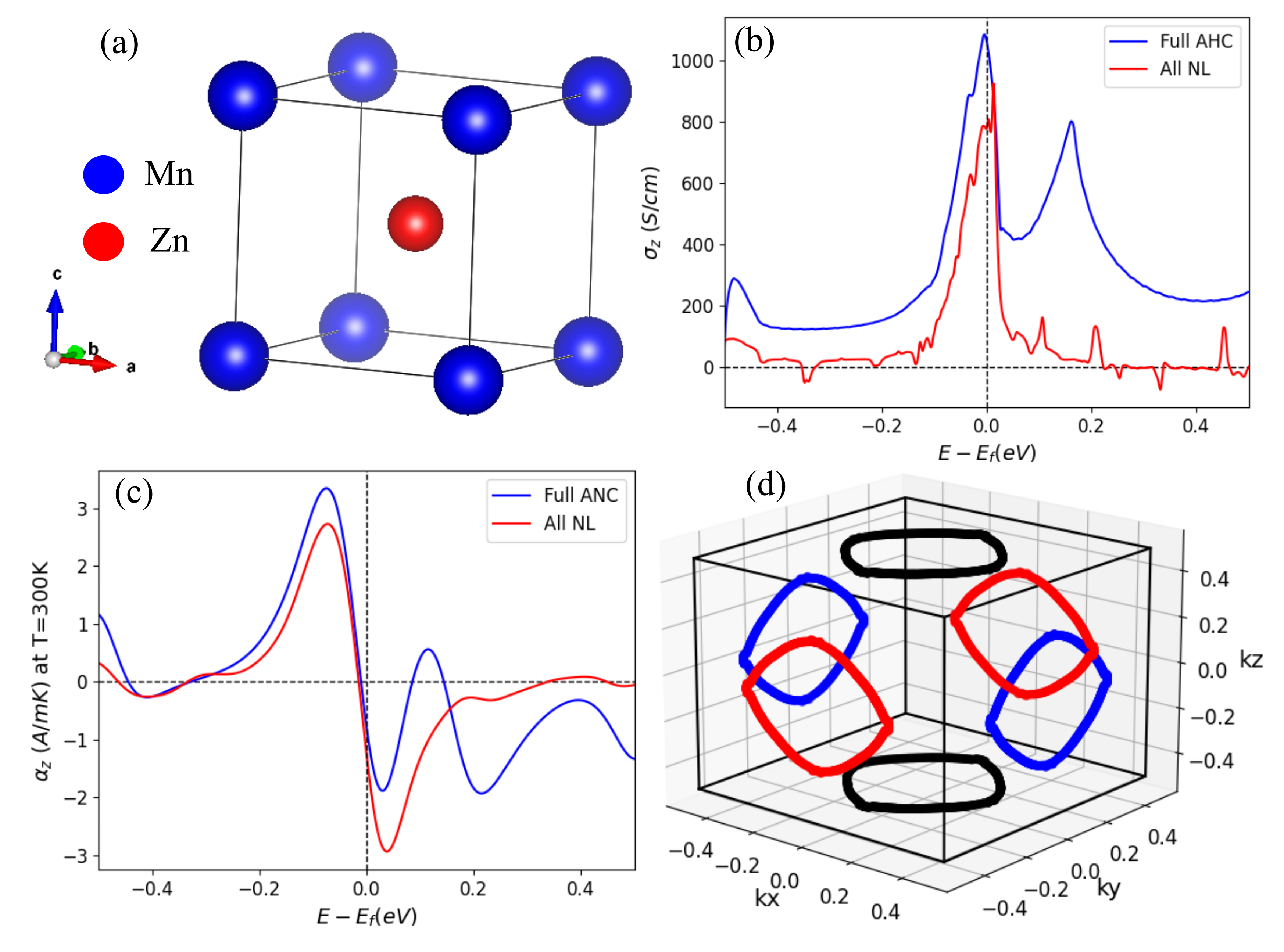}
\caption{(a) Crystal structure of MnZn. (b) The z-component of the AHC ($\sigma_z$) (blue) and the AHC contribution originating from the nodal lines (red) as a function of energy for MnZn with magnetization direction parallel to [001] axis. (c) The z-component of the ANC ($\alpha_z$) evaluated at T=300K (blue) and the ANC contribution originating from the nodal lines (red) as a function of energy for MnZn with magnetization direction parallel to [001] axis. (d) Symmetry related nodal lines of MnZn, with magnetiation direction parallel to [001] axis, contributing to the total AHC value.}
\label{fig2}
\end{figure*}

Recent works have shown that Weyl nodes and nodal lines, being bands touching points, behave either as sinks or as sources of the Berry curvature~\cite{nagaosa2010anomalous,singh2020multifunctional,xu2020high} and hence they are expected to contribute to the total AHC, the origin of which demonstrates the validity of the Mott relation. To further elucidate the origin of singular-like AHC with resulting magnificent ANC, detailed analysis is done on the energy-dependent AHC for MnZn (BNS: 123.345) (for crystal structure see Fig.~\ref{fig2}(a)), as shown in Fig.~\ref{fig2}(b) with the magnetization direction along the [001] axis. There exists a sharp peak of AHC about 1082$S/cm$ located about 4$meV$ below the Fermi energy. Explicit band structure analysis reveals the presence of a circularly shaped nodal line at $k_y=0.5$ plane (red part in Fig.~\ref{fig2}(d)), which further appears at the planes $k_x=\pm0.5$, $k_z=\pm0.5$ and $k_y=-0.5$ due to the presence of the symmetries of the compound (black and blue parts). Importantly, we found that about 73\% of the total AHC (782$S/cm$ out of 1082$S/cm$) can be attributed to the contribution of the nodal lines (Fig.~\ref{fig2}(b)), similarly to the role of Weyl nodes in Mn$_3$PdN~\cite{singh2020multifunctional}. Exactly due to the presence of such nodal lines and their contribution to the singular-like AHC, the behaviour of ANC around E$_F$ is dominated by the contributions from the nodal lines (Fig.~\ref{fig2}(c)). Therefore, the presence of Weyl nodes and nodal lines close to the Fermi energy can lead to an anomalous energy dependence of AHC and hence enhanced ANC.

A note about the role of symmetry of the nodal lines and their contributions to AHC is in order. As discussed above, the symmetry of the Berry Curvature in the Brillouin zone is essential to understand the origin of AHC. For MnZn with the magnetization direction along the z-axis, the corresponding Magnetic Laue group $4/mm'm'$ indicates the presence of the six closed-loop-nodal lines of Fig.~\ref{fig2}(d). However, not all of these symmetry equivalent nodal lines contribute equally to the same component of AHC, though their geometry are dictated by the energy eigenvalues and thus the symmetry. The underlined Laue group for the compound includes the $m_z$ mirror plane symmetry that flips the sign of $x$ and $y$ components of Berry curvature while it leaves $z$ component unchanged, according to:
\begin{align}
& \Omega_x\left(k_x,k_y,-k_z\right)=-\Omega_x\left(k_x,k_y,k_z\right) \notag \\
& \Omega_y\left(k_x,k_y,-k_z\right)=-\Omega_y\left(k_x,k_y,k_z\right) \notag \\
& \Omega_z\left(k_x,k_y,-k_z\right)=\Omega_z\left(k_x,k_y,k_z\right) \notag.
\end{align}
Hence, the contribution from the top black nodal ring as well as from the upper half of the red and the blue will exactly cancel out the one from the bottom black and the lower half of the red and the blue respectively for $x$ and $y$ components. A complete list of Berry curvature transformations with respect to each symmetry operation of the magnetic Laue group $4/mm'm'$  is found in Table S2 of the supplementary~\cite{supplementary}.

\begin{figure*}
\includegraphics[width=0.8\textwidth]{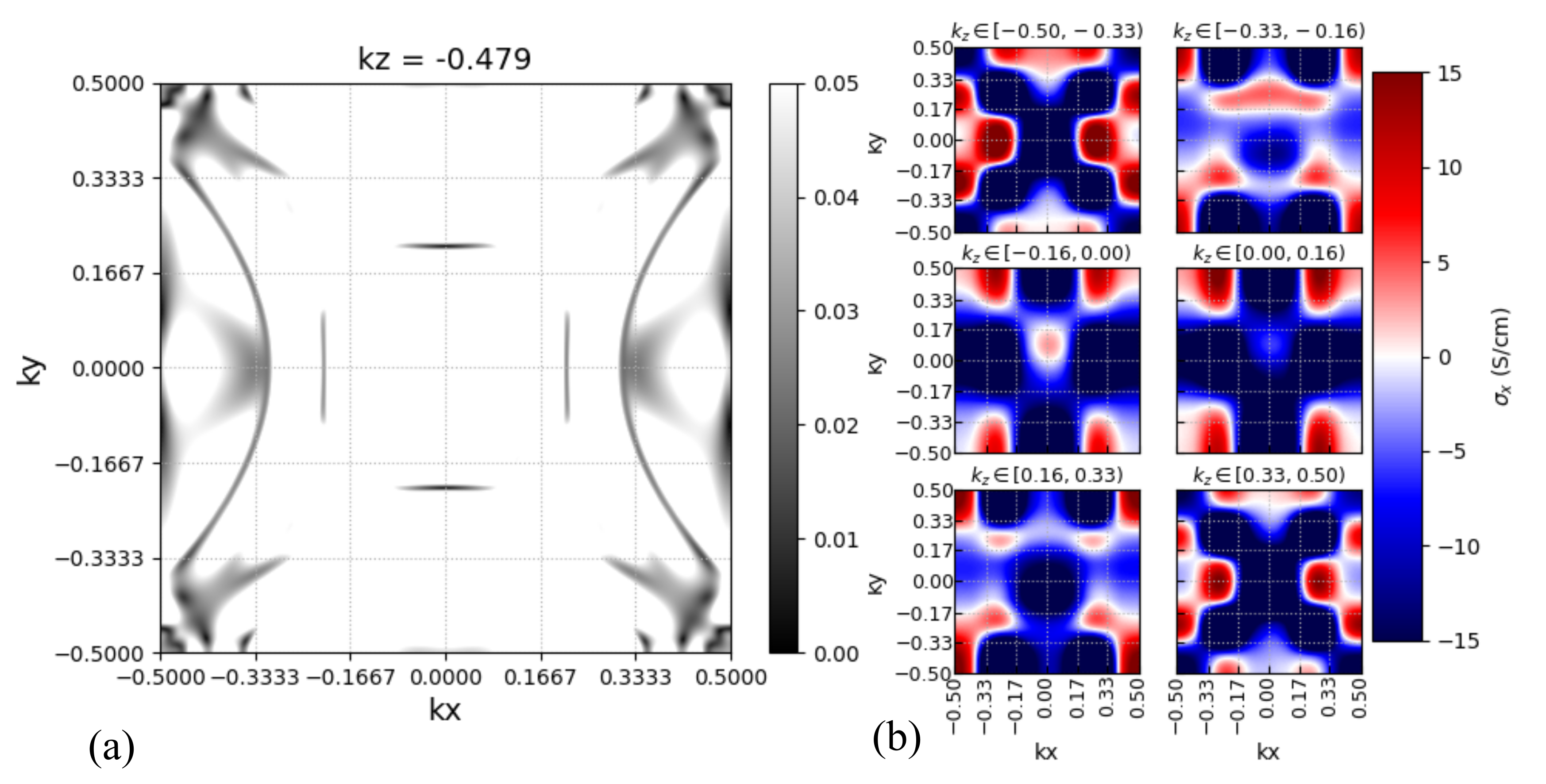}
\caption{(a) Band gap (in eV) at a slice of Brillouin Zone ($k_z=-0.479$) for Ni$_3$Pt. (b) The x-component of the AHC evaluated in 216 Brillouin Zone cubes for Ni$_3$Pt with magnetization direction parallel to [100] axis.}
\label{fig4}
\end{figure*}

Enhanced ANC values may originate from AHC contributions other than isolated Weyl nodes and nodal lines. The presence of nodal rings are responsible for almost 3/4 of the total AHC of MnZn, as discussed above. However, there are cases with more complicated behaviour, such as Ni$_3$Pt that hosts a surprisingly large AHC of -2051$S/cm$ that renders it as the largest calculated AHC. In order to investigate its origin, we split the BZ ($k_i \in \left[-0.500,0.500\right)$) in $6\times 6 \times 6 = 216$ cubes and calculate the AHC within each of these, as illustrated in Fig.~\ref{fig4}(b)). Our results show that each part of the BZ contributes to the total AHC, demonstrating the presence of numerous Weyl nodes, nodal lines as well as extended small band gap areas. Such areas are illustrated in the region $\left(k_x,k_y\right)\in\left[0.33,0.50\right]$ of the band gap plot of Fig.~\ref{fig4}(a). Despite the absence of isolated Weyl nodes, there is a giant ANC of -7.29$A/(m\cdot K)$ (at the Fermi energy) calculated, originating from the whole BZ, showing at the same time that large ANC values are possible even in the absence of isolated hot-spot contributions to the Berry curvature.   

\begin{figure*}
\includegraphics[width=0.8\textwidth]{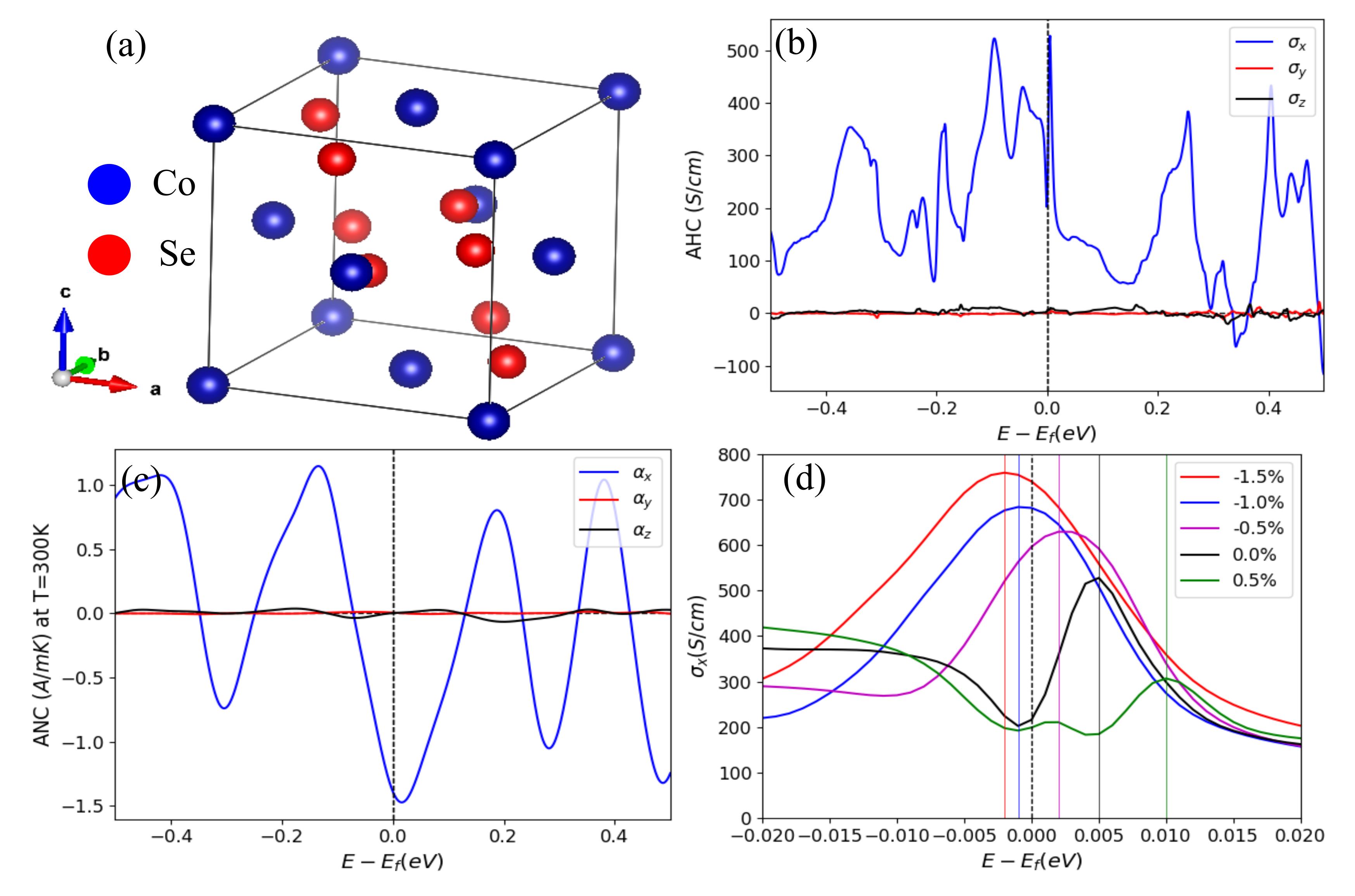}
\caption{(a) Crystal structure of CoSe$_4$. (b) The AHC components as a function of energy for CoSe$_4$ with magnetization direction parallel to [100] axis. (c) The ANC components evaluated at T=300K as a function of energy for CoSe$_4$ with magnetization direction parallel to [100] axis. (d) The x-component of the AHC of CoSe$_4$ with magnetization direction parallel to [100] axis for different values of applied strain}
\label{fig3}
\end{figure*}

As the Fermi level can be tuned by several mechanisms including doping and mechanical strain, it is fruitful to discuss the possibility of inducing larger AHC and ANC values away from the charge neutral point. In order to investigate the impact of such modifications to the transport properties, we consider the hexagonal compound CoSe$_4$ (BNS: 61.436) (for crystal structure see Fig.~\ref{fig3}(a). Interestingly, the compound exhibits an AHC of 202$S/cm$ at the Fermi energy and moreover a sharp peak at almost 500$S/cm$, 5 meV above (see  Fig.~\ref{fig3}(b)) that can be tuned by doping. Since, in general, compounds that exhibit single-valued AHC peaks close to the Fermi energy offer great opportunities for modifying AHC and and hence ANC values by means of doping, some promising candidates include VAu$_4$, CoS$_2$ and Co$_2$CrAl that exhibit AHC peaks of 1955$S/cm$, -1048$S/cm$ and 1140$S/cm$ at 0.051$eV$, -0.024$eV$ and 0.027$eV$ respectively. AHC and ANC can also be modified by applying biaxial strain. That is, the tensile strain of 0.5\% reduces the size of the peak to 306$S/cm$ and moves it at 0.01$eV$ in respect to the Fermi energy. On the other hand, the applied compressive strain of 1.0\% has the opposite effect and hence the size of the peak is increased to 683$S/cm$ and its position lies at -0.001$eV$ (see Fig.~\ref{fig3}(d)).

\section{\label{CONC}Conclusion}

Based on high throughput first-principles calculations, we evaluated the anomalous Hall conductivity as well as the anomalous Nernst conductivity of 266 transition-metal based ferromagnets. We report that the absolute value of AHC (ANC) of 11 (16) compounds is larger than 1000$S/cm$ (3$A/(m\cdot K)$) with the largest being equal to 2060$S/cm$ for CrPt$_3$ (-7.24$A/(m\cdot K)$ for Ni$_3$Pt). Moreover, we find that the AHC and ANC values are by 3/4 originating from linear degenerate states such as Weyl nodes and nodal lines in MnZn and they can further be enhanced by a factor of 43\% by applying external stimuli, such as 1.5\% compressive strain in CoSe$_4$. However, large AHC and ANC values can arise even without the presence of isolated singular hot spots in the Berry curvature but instead from uniform extended small band gap areas from the whole BZ in Ni$_3$Pt.
 
\begin{center}
\small \textbf{ACKNOWLEDGMENTS}
\end{center}

This work was financially supported by the Deutsche Forschungsgemeinschaft (DFG) via the priority programme SPP 1666 and the calculations were conducted on the Lichtenberg high performance computer of the TU Darmstadt.

\end{document}


\maketitle

\makeatletter
\def\highlight#1{%
\fboxrule2pt %
\hsize=\dimexpr\hsize-2\fboxrule-2\fboxsep\relax
#1%
\@endpbox\unskip\setbox0\lastbox\bgroup
\fboxrule2pt %
\fcolorbox{red}{lightgray}{\box0}\hfill}

\vspace{-2cm}

\begin{tiny}
\begin{longtable}{|c|c|c|c|c|c|c|c|c|c|c|}
\hline
\bf N & \bf Compound & \bf SPG & $\mathbf{\sigma_x}$ & $\mathbf{\alpha_x}$ & $\mathbf{\sigma_z}$ & $\mathbf{\alpha_z}$ & \bf Literature & \bf Literature & \bf Anisotropy & \bf Anisotropy \\ 
  &   &   & $\left[\frac{S}{cm}\right]$ & $\left[\frac{A}{m \cdot K}\right]$ & $\left[\frac{S}{cm}\right]$ & $\left[\frac{A}{m \cdot K}\right]$ & $\mathbf{\sigma_z} \left[\frac{S}{cm}\right]$ & $\mathbf{\alpha_z} \left[\frac{A}{m \cdot K}\right]$ & \bf in AHC & \bf in ANC \\ 
\hline 
\hline 
1 & Cr$_{3}$Te$_{4}$ & 12 & 27.58 & 1.06 & 315.96 & 0.52 & - & 0.60 at 500K~\cite{sakai2020iron} & \cellcolor{green}0.09 & 2.05 \\ 
\rowcolor{Gray}
2 & Te$_{16}$Cr$_{14}$ & 12 & 441.73 & -1.29 & 291.11 & -1.00 & - & - & 1.52 & 1.28 \\ 
3 & Se$_{16}$Cr$_{14}$ & 12 & 240.75 & 0.22 & 146.16 & -1.40 & - & - & 1.65 & \cellcolor{green}-0.16 \\ 
\rowcolor{Gray}
4 & Fe$_{3}$Se$_{4}$ & 12 & 114.49 & -1.47 & 469.53 & 0.04 & - & 1.48 at 500K~\cite{sakai2020iron} & \cellcolor{green}0.24 & \cellcolor{green}-33.39 \\ 
5 & Cr$_{3}$Se$_{4}$ & 12 & 301.94 & 0.87 & -36.22 & 0.36 & - & 0.58 at 500K~\cite{sakai2020iron} & \cellcolor{green}-8.34 & 2.46 \\ 
\rowcolor{Gray}
6 & Hf$_{2}$Co$_{7}$ & 12 & 817.75 & 0.37 & 402.61 & -0.14 & - & - & 2.03 & -2.65 \\ 
7 & Mn$_{6}$As$_{4}$ & 12 & 820.61 & -0.03 & 373.57 & 2.46 & - & - & 2.20 & \cellcolor{green}-0.01 \\ 
\rowcolor{Gray}
8 & K$_{4}$Mn$_{4}$F$_{16}$ & 14 & 0.00 & 0.00 & -0.00 & -0.00 & - & - & -0.54 & -0.24 \\ 
9 & C$_{4}$Fe$_{10}$ & 15 & 703.86 & -1.96 & 554.15 & 0.56 & - & - & 1.27 & -3.48 \\ 
\rowcolor{Gray}
10 & Ba$_{2}$V$_{2}$S$_{6}$ & 20 & 67.11 & 0.29 & -12.88 & 0.12 & - & - & \cellcolor{green}-5.21 & 2.34 \\ 
11 & Se$_{8}$Ag$_{4}$Ge$_{2}$Mn$_{2}$ & 31 & -0.61 & -0.02 & 7.70 & -0.11 & - & - & -0.08 & \cellcolor{green}0.15 \\ 
\rowcolor{Gray}
12 & Co$_{2}$Nb$_{1}$Sn$_{1}$ & 51 & -15.15 & -0.59 & 91.57 & 0.17 & - & 1.76 at 500K~\cite{sakai2020iron} & \cellcolor{green}-0.17 & -3.41 \\ 
13 & C$_{8}$N$_{12}$Co$_{2}$ & 58 & 18.70 & -0.60 & 20.68 & 0.19 & - & - & 0.90 & -3.13 \\ 
\rowcolor{Gray}
14 & C$_{8}$N$_{12}$Ni$_{2}$ & 58 & -0.06 & -0.04 & 0.01 & 0.00 & - & - & -9.30 & -14.21 \\ 
15 & P$_{4}$Mn$_{4}$ & 62 & 154.64 & -0.63 & 160.07 & -1.58 & - & - & 0.97 & 0.40 \\ 
\rowcolor{Gray}
16 & Mn$_{4}$Co$_{4}$Ge$_{4}$ & 62 & 155.18 & -0.85 & 113.56 & -0.27 & - & - & 1.37 & 3.19 \\ 
17 & Mn$_{4}$Co$_{4}$Si$_{4}$ & 62 & 6.31 & 0.79 & -14.53 & 0.74 & - & - & -0.43 & 1.07 \\ 
\rowcolor{Gray}
18 & Fe$_{12}$C$_{4}$ & 62 & 432.26 & 0.82 & 690.77 & 0.67 & - & - & 0.63 & 1.22 \\ 
19 & Mn$_{4}$Co$_{4}$P$_{4}$ & 62 & 408.78 & 0.39 & -45.07 & 1.11 & - & - & \cellcolor{green}-9.07 & 0.35 \\ 
\rowcolor{Gray}
20 & Si$_{4}$Mn$_{4}$Ni$_{4}$ & 62 & 116.73 & 0.43 & 451.49 & 0.61 & - & - & 0.26 & 0.71 \\ 
21 & Cr$_{4}$Ni$_{4}$P$_{4}$ & 62 & 78.51 & -0.24 & 237.27 & 0.14 & - & - & 0.33 & -1.64 \\ 
\rowcolor{Gray}
22 & B$_{4}$Fe$_{4}$ & 62 & 78.15 & 2.12 & 31.24 & 0.63 & - & - & 2.50 & 3.38 \\ 
23 & Fe$_{4}$Co$_{4}$P$_{4}$ & 62 & 329.66 & 0.72 & 26.17 & -0.08 & - & - & \cellcolor{green}12.60 & \cellcolor{green}-9.20 \\ 
\rowcolor{Gray}
24 & P$_{4}$Fe$_{4}$ & 62 & 0.46 & 0.16 & -48.42 & -0.38 & - & - & \cellcolor{green}-0.01 & -0.42 \\ 
25 & Fe$_{12}$B$_{4}$ & 62 & 924.23 & -2.96 & 522.07 & -1.72 & - & - & 1.77 & 1.72 \\ 
\rowcolor{Gray}
26 & Si$_{4}$S$_{16}$Mn$_{8}$ & 62 & 0.01 & -0.04 & 0.03 & -0.14 & - & - & 0.31 & 0.31 \\ 
27 & F$_{8}$K$_{4}$Cu$_{2}$ & 64 & 0.06 & -0.00 & 0.05 & -0.00 & - & - & 1.13 & 0.32 \\ 
\rowcolor{Gray}
28 & Mn$_{3}$Pd$_{5}$ & 65 & 223.49 & 0.19 & 155.85 & 0.17 & - & - & 1.43 & 1.12 \\ 
29 & Mn$_{10}$Ge$_{4}$ & 72 & 791.05 & 1.77 & 185.03 & 1.52 & - & - & \cellcolor{green}4.28 & 1.17 \\ 
\rowcolor{Gray}
30 & P$_{4}$Fe$_{12}$ & 82 & 238.85 & -1.39 & 276.83 & -1.63 & -54~\cite{noky2020giant} & -1.32 at 300K~\cite{noky2020giant} & 0.86 & 0.85 \\ 
31 & Ni$_{12}$P$_{4}$ & 82 & 12.64 & 0.24 & -7.57 & -0.02 & - & - & -1.67 & \cellcolor{green}-10.04 \\ 
\rowcolor{Gray}
32 & Mn$_{1}$Au$_{4}$ & 87 & 272.27 & 0.24 & 154.49 & 0.40 & - & 0.33 at 500K~\cite{sakai2020iron} & 1.76 & 0.61 \\ 
33 & V$_{1}$Au$_{4}$ & 87 & \cellcolor{cyan}1152.27 & \cellcolor{cyan}3.13 & \cellcolor{cyan}1036.60 & 1.11 & - & - & 1.11 & 2.83 \\ 
\rowcolor{Gray}
34 & Cu$_{4}$Fe$_{4}$Se$_{8}$ & 112 & 515.61 & -0.77 & 761.76 & \cellcolor{cyan}-3.62 & - & - & 0.68 & \cellcolor{green}0.21 \\ 
35 & Fe$_{1}$Pt$_{1}$ & 123 & 440.96 & -0.71 & 861.79 & -0.60 &  818~\cite{weischenberg2011ab} & 1.35 at 500K~\cite{sakai2020iron} & 0.51 & 1.18 \\ 
\rowcolor{Gray}
36 & Pt$_{1}$Cr$_{1}$ & 123 & 403.08 & -0.29 & -105.21 & 0.91 & - & - & -3.83 & -0.32 \\ 
37 & Fe$_{2}$Ni$_{2}$ & 123 & 313.37 & 0.01 & 153.52 & -0.29 & - & - & 2.04 & \cellcolor{green}-0.05 \\ 
\rowcolor{Gray}
38 & Co$_{2}$Pt$_{2}$ & 123 & 233.05 & \cellcolor{cyan}-4.69 & 79.55 & -2.35 & - & - & 2.93 & 2.00 \\ 
39 & Fe$_{2}$Pd$_{2}$ & 123 & 122.29 & -0.09 & 290.13 & -0.21 & 133~\cite{weischenberg2011ab} & - & 0.42 & 0.42 \\ 
\rowcolor{Gray}
40 & Mn$_{4}$Ga$_{10}$ & 127 & 294.98 & 0.09 & 345.73 & 1.04 & - & - & 0.85 & \cellcolor{green}0.08 \\ 
41 & Rh$_{10}$Sc$_{4}$B$_{4}$Fe$_{2}$ & 127 & -14.85 & 1.23 & 83.92 & 0.05 & - & - & \cellcolor{green}-0.18 & \cellcolor{green}22.55 \\ 
\rowcolor{Gray}
42 & Mn$_{4}$Sb$_{2}$ & 129 & 440.31 & 2.25 & 445.24 & 1.86 & - & - & 0.99 & 1.21 \\ 
43 & Al$_{2}$Mn$_{2}$Ge$_{2}$ & 129 & 121.35 & 0.69 & 62.23 & 0.18 & - & - & 1.95 & 3.80 \\ 
\rowcolor{Gray}
44 & Si$_{2}$Mn$_{2}$Y$_{2}$ & 129 & -20.19 & -0.84 & 223.91 & -0.16 & - & - & \cellcolor{green}-0.09 & \cellcolor{green}5.11 \\ 
45 & Al$_{8}$Fe$_{4}$Y$_{1}$ & 139 & 36.33 & \cellcolor{cyan}-3.13 & 327.58 & -0.11 & - & - & \cellcolor{green}0.11 & \cellcolor{green}27.92 \\ 
\rowcolor{Gray}
46 & Y$_{2}$Fe$_{24}$N$_{2}$ & 139 & 106.23 & \cellcolor{cyan}3.45 & 29.26 & 0.76 & - & - & 3.63 & \cellcolor{green}4.56 \\ 
47 & Tl$_{2}$Cu$_{4}$Se$_{4}$ & 139 & -4.99 & -0.03 & -1.97 & -0.04 & - & - & 2.54 & 0.65 \\ 
\rowcolor{Gray}
48 & Rb$_{2}$Cr$_{1}$Cl$_{4}$ & 139 & 0.15 & -0.00 & -7.57 & -0.05 & - & 0.90 at 500K~\cite{sakai2020iron} & -0.02 & 0.00 \\ 
49 & S$_{2}$Co$_{2}$Tl$_{1}$ & 139 & 43.58 & -0.27 & -200.52 & -0.27 & - & 0.78 at 500K~\cite{sakai2020iron} & \cellcolor{green}-0.22 & 1.01 \\ 
\rowcolor{Gray}
50 & Fe$_{8}$W$_{8}$Fe$_{8}$Y$_{2}$ & 139 & 936.06 & -0.72 & 316.07 & -1.97 & - & - & 2.96 & 0.37 \\ 
51 & Co$_{1}$Sn$_{1}$Rh$_{2}$ & 139 & 311.28 & 1.45 & 337.86 & 1.13 & - & - & 0.92 & 1.28 \\ 
\rowcolor{Gray}
52 & Fe$_{1}$Sn$_{1}$Rh$_{2}$ & 139 & -204.19 & -1.23 & 436.68 & 0.48 & - & - & -0.47 & -2.58 \\ 
53 & V$_{1}$Pt$_{3}$ & 139 & 314.51 & -0.31 & -292.64 & 1.27 & 900~\cite{markou2020hard} & 7.02 at 500K~\cite{sakai2020iron} & -1.07 & \cellcolor{green}-0.25 \\ 
\rowcolor{Gray}
54 & Mn$_{2}$Ge$_{2}$Y$_{1}$ & 139 & 83.68 & 0.27 & 266.29 & 0.25 & - & 0.05 at 500K~\cite{sakai2020iron} & 0.31 & 1.10 \\ 
55 & Be$_{12}$Cr$_{1}$ & 139 & 25.24 & 0.38 & 24.54 & 0.07 & - & - & 1.03 & \cellcolor{green}5.36 \\ 
\rowcolor{Gray}
56 & Rh$_{4}$Co$_{2}$Sb$_{2}$ & 139 & 217.67 & -0.31 & 87.49 & -0.92 & - & - & 2.49 & 0.34 \\ 
57 & Rh$_{4}$Fe$_{2}$Sb$_{2}$ & 139 & 124.62 & -0.32 & 81.72 & -1.10 & - & - & 1.53 & 0.29 \\ 
\rowcolor{Gray}
58 & K$_{1}$Co$_{2}$Se$_{2}$ & 139 & 168.66 & -0.08 & -180.69 & -0.28 & - & 0.56 at 500K~\cite{sakai2020iron} & -0.93 & 0.29 \\ 
59 & Mn$_{1}$Sb$_{1}$Rh$_{2}$ & 139 & 245.36 & -0.52 & 4.60 & 0.73 & 4~\cite{noky2020giant} & 2.62 at 300K~\cite{noky2020giant} & \cellcolor{green}53.32 & -0.71 \\ 
\rowcolor{Gray}
60 & Rb$_{2}$Cu$_{1}$F$_{4}$ & 139 & -1.35 & 0.00 & -1.12 & 0.00 & - & - & 1.21 & 0.45 \\ 
61 & Se$_{4}$Co$_{4}$Tl$_{2}$ & 139 & 106.36 & 0.00 & -148.57 & -1.33 & - & - & -0.72 & \cellcolor{green}-0.00 \\ 
\rowcolor{Gray}
62 & Fe$_{2}$Ge$_{4}$ & 140 & 256.17 & 0.02 & 304.81 & 0.39 & - & - & 0.84 & \cellcolor{green}0.04 \\ 
63 & Co$_{4}$B$_{2}$ & 140 & 703.07 & -0.15 & 581.86 & 1.23 & - & - & 1.21 & \cellcolor{green}-0.12 \\ 
\rowcolor{Gray}
64 & Fe$_{4}$B$_{2}$ & 140 & 414.57 & -0.84 & 483.77 & -1.49 & - & - & 0.86 & 0.57 \\ 
65 & Te$_{6}$Cr$_{2}$Ge$_{2}$ & 148 & 0.00 & 0.02 & -0.00 & -0.02 & - & - & -0.49 & -0.77 \\ 
\rowcolor{Gray}
66 & Si$_{2}$Cr$_{2}$Te$_{6}$ & 148 & -0.00 & -0.02 & -0.00 & -0.01 & - & - & 1.03 & 2.65 \\ 
67 & F$_{6}$V$_{1}$Nb$_{1}$ & 148 & -2.75 & -0.07 & 24.14 & 0.75 & - & 0.05 at 500K~\cite{sakai2020iron} & \cellcolor{green}-0.11 & \cellcolor{green}-0.09 \\ 
\rowcolor{Gray}
68 & Ni$_{1}$Pt$_{1}$F$_{6}$ & 148 & -0.10 & -0.00 & -0.06 & 0.03 & - & 0.23 at 500K~\cite{sakai2020iron} & 1.72 & -0.07 \\ 
69 & F$_{6}$Pd$_{1}$Pt$_{1}$ & 148 & 0.01 & -0.11 & 6.49 & -1.04 & - & 0.48 at 500K~\cite{sakai2020iron} & 0.00 & \cellcolor{green}0.10 \\ 
\rowcolor{Gray}
70 & Mn$_{1}$Zn$_{1}$F$_{6}$ & 148 & 0.00 & -0.00 & -0.04 & 0.03 & - & - & -0.00 & -0.00 \\ 
71 & S$_{8}$V$_{4}$Ga$_{1}$ & 160 & 4.02 & -0.02 & 3.93 & -0.03 & - & - & 1.02 & 0.70 \\ 
\rowcolor{Gray}
72 & Ga$_{1}$Mo$_{4}$S$_{8}$ & 160 & 0.05 & 0.01 & 0.37 & -0.02 & - & - & 0.12 & -0.66 \\ 
73 & Ga$_{1}$Mo$_{4}$Se$_{8}$ & 160 & -0.07 & -0.00 & 0.26 & -0.01 & - & - & -0.28 & 0.17 \\ 
\rowcolor{Gray}
74 & N$_{3}$Fe$_{6}$ & 162 & 436.08 & 0.11 & 190.10 & 0.88 & - & - & 2.29 & \cellcolor{green}0.12 \\ 
75 & Cr$_{8}$Te$_{12}$ & 163 & -163.44 & -0.98 & -18.91 & 0.56 & - & - & \cellcolor{green}8.64 & -1.76 \\ 
\rowcolor{Gray}
76 & Cr$_{10}$S$_{12}$ & 163 & -79.64 & 0.05 & -70.59 & -0.01 & - & - & 1.13 & \cellcolor{green}-6.47 \\ 
77 & S$_{2}$Ta$_{1}$ & 164 & -3.96 & -0.00 & -9.27 & -0.12 & - & - & 0.43 & \cellcolor{green}0.02 \\ 
\rowcolor{Gray}
78 & Y$_{3}$Fe$_{9}$ & 166 & 304.96 & -0.64 & 64.59 & 1.92 & - & - & \cellcolor{green}4.72 & -0.33 \\ 
79 & Co$_{3}$Sn$_{2}$S$_{2}$ & 166 & 757.13 & 1.01 & 72.17 & 1.64 & 1130~\cite{liu2018giant} & 1.35 at 500K~\cite{sakai2020iron} & \cellcolor{green}10.49 & 0.62 \\ 
\rowcolor{Gray}
80 & Tl$_{2}$Fe$_{6}$Te$_{6}$ & 176 & -252.63 & -0.17 & -77.75 & -0.30 & - & - & 3.25 & 0.57 \\ 
81 & Se$_{16}$Nb$_{8}$Mn$_{2}$ & 176 & -34.74 & -0.81 & 216.66 & 2.57 & - & - & \cellcolor{green}-0.16 & -0.32 \\ 
\rowcolor{Gray}
82 & N$_{2}$Fe$_{6}$ & 182 & 234.04 & 0.27 & 111.88 & 0.21 & - & - & 2.09 & 1.31 \\ 
83 & Cr$_{2}$Nb$_{6}$S$_{12}$ & 182 & -2.26 & -0.14 & -53.97 & -0.15 & - & - & \cellcolor{green}0.04 & 0.97 \\ 
\rowcolor{Gray}
84 & S$_{12}$Mn$_{2}$Nb$_{6}$ & 182 & -12.32 & 0.18 & -64.10 & 0.07 & - & - & \cellcolor{green}0.19 & 2.67 \\ 
85 & Se$_{12}$Ta$_{4}$Cr$_{2}$Ta$_{2}$ & 182 & -137.59 & 0.07 & -195.55 & -0.22 & - & - & 0.70 & -0.34 \\ 
\rowcolor{Gray}
86 & Fe$_{2}$Ta$_{6}$S$_{12}$ & 182 & -190.51 & -0.35 & -363.98 & 0.36 & - & - & 0.52 & -0.98 \\ 
87 & S$_{2}$Cd$_{2}$ & 186 & -2.39 & 0.00 & -0.21 & 0.01 & - & - & 11.17 & 0.27 \\ 
\rowcolor{Gray}
88 & Se$_{2}$Cd$_{2}$ & 186 & 0.00 & 0.03 & 0.00 & 0.03 & - & - & 1.28 & 0.98 \\ 
89 & P$_{3}$Fe$_{6}$ & 189 & 217.79 & -0.86 & -51.95 & -0.71 & - & - & \cellcolor{green}-4.19 & 1.21 \\ 
\rowcolor{Gray}
90 & Mn$_{3}$As$_{3}$Rh$_{3}$ & 189 & 442.21 & -1.04 & -143.21 & 1.27 & - & - & -3.09 & -0.82 \\ 
91 & Cr$_{3}$Ni$_{3}$As$_{3}$ & 189 & 89.23 & -0.36 & 270.38 & 0.62 & - & - & 0.33 & -0.58 \\ 
\rowcolor{Gray}
92 & Mn$_{3}$P$_{3}$Rh$_{3}$ & 189 & 202.71 & 0.26 & -120.82 & -0.30 & - & - & -1.68 & -0.89 \\ 
93 & Pd$_{3}$Cr$_{3}$As$_{3}$ & 189 & 78.81 & -0.43 & 163.64 & \cellcolor{cyan}-3.66 & - & - & 0.48 & \cellcolor{green}0.12 \\ 
\rowcolor{Gray}
94 & Mn$_{3}$Fe$_{3}$As$_{3}$ & 189 & 33.99 & -0.92 & -177.30 & 0.15 & - & - & \cellcolor{green}-0.19 & \cellcolor{green}-6.21 \\ 
95 & Mn$_{3}$As$_{3}$Pd$_{3}$ & 189 & 148.17 & -0.91 & 236.86 & 0.68 & - & - & 0.63 & -1.34 \\ 
\rowcolor{Gray}
96 & Mn$_{3}$Ge$_{3}$Pd$_{3}$ & 189 & 463.46 & -0.39 & 420.74 & -0.98 & - & - & 1.10 & 0.40 \\ 
97 & Mn$_{3}$As$_{3}$Ru$_{3}$ & 189 & 499.62 & -2.28 & 226.23 & \cellcolor{cyan}-3.38 & - & - & 2.21 & 0.68 \\ 
\rowcolor{Gray}
98 & B$_{2}$Co$_{8}$Y$_{2}$ & 191 & 554.19 & -2.93 & -938.76 & \cellcolor{cyan}5.83 & - & - & -0.59 & -0.50 \\ 
99 & Y$_{1}$Co$_{5}$ & 191 & \cellcolor{cyan}1142.26 & \cellcolor{cyan}-3.84 & 256.80 & \cellcolor{cyan}3.33 & - & 6.05 at 500K~\cite{sakai2020iron} & \cellcolor{green}4.45 & -1.15 \\ 
\rowcolor{Gray}
100 & Fe$_{11}$B$_{4}$Y$_{3}$ & 191 & 269.20 & 0.73 & 919.22 & -2.81 & - & - & 0.29 & -0.26 \\ 
101 & Y$_{2}$Fe$_{2}$Co$_{6}$B$_{2}$ & 191 & 34.83 & 0.63 & -385.88 & 1.18 & - & - & \cellcolor{green}-0.09 & 0.53 \\ 
\rowcolor{Gray}
102 & Y$_{3}$Co$_{11}$B$_{4}$ & 191 & 997.23 & -0.45 & -201.09 & 1.77 & - & - & \cellcolor{green}-4.96 & -0.26 \\ 
103 & Y$_{4}$Co$_{14}$B$_{6}$ & 191 & 716.92 & 2.29 & -28.26 & 0.11 & - & - & \cellcolor{green}-25.37 & \cellcolor{green}20.85 \\ 
\rowcolor{Gray}
104 & Ni$_{13}$Y$_{3}$B$_{2}$ & 191 & 403.96 & -0.67 & -264.76 & -2.65 & - & - & -1.53 & 0.25 \\ 
105 & Mn$_{6}$Sn$_{6}$Mg$_{1}$ & 191 & 175.79 & 0.21 & -6.25 & -0.82 & - & - & \cellcolor{green}-28.14 & -0.26 \\ 
\rowcolor{Gray}
106 & Mn$_{1}$B$_{2}$ & 191 & 66.43 & 0.85 & -107.00 & 0.58 & - & 1.19 at 500K~\cite{sakai2020iron} & -0.62 & 1.46 \\ 
107 & Y$_{1}$Ni$_{5}$ & 191 & -246.34 & -0.73 & 113.57 & 0.25 & - & 1.82 at 500K~\cite{sakai2020iron} & -2.17 & -2.87 \\ 
\rowcolor{Gray}
108 & Zr$_{1}$Mn$_{6}$Sn$_{6}$ & 191 & 993.71 & 1.29 & 346.42 & -0.82 & - & - & 2.87 & -1.57 \\ 
109 & Mn$_{10}$Ge$_{6}$ & 193 & 196.23 & -1.18 & 964.93 & 1.69 & - & - & \cellcolor{green}0.20 & -0.70 \\ 
\rowcolor{Gray}
110 & Si$_{6}$Fe$_{10}$ & 193 & 209.47 & -0.57 & -8.44 & -2.41 & - & - & \cellcolor{green}-24.82 & \cellcolor{green}0.24 \\ 
111 & Cr$_{4}$Te$_{4}$ & 194 & -521.03 & 0.44 & -513.33 & 0.58 & - & - & 1.01 & 0.76 \\ 
\rowcolor{Gray}
112 & Y$_{2}$Fe$_{17}$ & 194 & 374.01 & -0.63 & 474.77 & -0.96 & - & - & 0.79 & 0.66 \\ 
113 & Fe$_{6}$Ge$_{2}$ & 194 & 417.80 & -0.85 & 197.21 & 0.29 & - & - & 2.12 & -2.94 \\ 
\rowcolor{Gray}
114 & Mn$_{2}$As$_{2}$ & 194 & -38.70 & 0.02 & 178.43 & -1.63 & - & - & \cellcolor{green}-0.22 & \cellcolor{green}-0.02 \\ 
115 & Sc$_{4}$Fe$_{8}$ & 194 & 213.10 & -1.09 & 189.90 & 1.09 & - & - & 1.12 & -1.00 \\ 
\rowcolor{Gray}
116 & Sc$_{6}$In$_{2}$ & 194 & -6.66 & -0.75 & -139.79 & 0.04 & - & - & \cellcolor{green}0.05 & \cellcolor{green}-20.09 \\ 
117 & Hf$_{4}$Fe$_{8}$ & 194 & 284.54 & 0.45 & 415.10 & 2.42 & - & - & 0.69 & \cellcolor{green}0.19 \\ 
\rowcolor{Gray}
118 & Be$_{8}$Fe$_{4}$ & 194 & 475.24 & -1.00 & 385.15 & \cellcolor{cyan}-3.91 & - & - & 1.23 & 0.26 \\ 
119 & Co$_{2}$ & 194 & 38.95 & -0.13 & 482.78 & 0.05 & 477~\cite{weischenberg2011ab} & - & \cellcolor{green}0.08 & -2.39 \\ 
\rowcolor{Gray}
120 & F$_{6}$Ni$_{2}$Cs$_{2}$ & 194 & 0.06 & -0.02 & -0.00 & 0.00 & - & - & -3023.06 & -7987.00 \\ 
121 & Fe$_{2}$Co$_{2}$Ge$_{2}$ & 194 & -26.92 & -2.25 & 706.25 & 0.85 & - & - & \cellcolor{green}-0.04 & -2.65 \\ 
\rowcolor{Gray}
122 & Mn$_{2}$Bi$_{2}$ & 194 & \cellcolor{cyan}1599.55 & -1.42 & 849.95 & 2.35 & - & - & 1.88 & -0.60 \\ 
123 & Mn$_{2}$Co$_{2}$Sn$_{2}$ & 194 & 382.66 & \cellcolor{cyan}-3.65 & -280.30 & \cellcolor{cyan}-3.20 & - & - & -1.37 & 1.14 \\ 
\rowcolor{Gray}
124 & Mn$_{2}$Ga$_{2}$Pt$_{2}$ & 194 & \cellcolor{cyan}-1081.62 & \cellcolor{cyan}-3.63 & 611.59 & -0.97 & - & - & -1.77 & 3.75 \\ 
125 & Mn$_{6}$Ga$_{2}$ & 194 & 119.35 & 1.73 & \cellcolor{cyan}1605.63 & 3.91 & - & - & \cellcolor{green}0.07 & 0.44 \\ 
\rowcolor{Gray}
126 & Zr$_{4}$Fe$_{2}$V$_{6}$ & 194 & -27.76 & 1.02 & -205.58 & 0.28 & - & - & \cellcolor{green}0.14 & 3.59 \\ 
127 & Fe$_{6}$Sn$_{2}$ & 194 & 802.88 & -1.04 & 822.55 & \cellcolor{cyan}-3.13 & -503~\cite{noky2020giant} & 1.02 at 300K~\cite{noky2020giant} & 0.98 & 0.33 \\ 
\rowcolor{Gray}
128 & Mg$_{4}$Co$_{8}$ & 194 & -871.05 & 1.02 & 204.33 & 0.07 & - & - & \cellcolor{green}-4.26 & \cellcolor{green}14.89 \\ 
129 & Fe$_{4}$Ge$_{2}$ & 194 & -464.24 & -0.01 & \cellcolor{cyan}1182.00 & -2.35 & - & - & -0.39 & \cellcolor{green}0.00 \\ 
\rowcolor{Gray}
130 & Mn$_{2}$Ni$_{2}$Ge$_{2}$ & 194 & 213.35 & 0.65 & 90.35 & 1.63 & - & - & 2.36 & 0.40 \\ 
131 & Nb$_{4}$Fe$_{8}$ & 194 & \cellcolor{cyan}1212.37 & \cellcolor{cyan}4.01 & \cellcolor{cyan}1029.89 & -1.23 & - & - & 1.18 & -3.25 \\ 
\rowcolor{Gray}
132 & Cl$_{6}$Rb$_{2}$Fe$_{2}$ & 194 & -4.69 & -0.66 & 0.81 & 0.05 & - & - & -5.81 & \cellcolor{green}-13.57 \\ 
133 & Ti$_{4}$Fe$_{8}$ & 194 & 821.15 & \cellcolor{cyan}3.21 & 96.72 & \cellcolor{cyan}5.68 & - & - & \cellcolor{green}8.49 & 0.56 \\ 
\rowcolor{Gray}
134 & Al$_{6}$Fe$_{2}$Y$_{4}$ & 194 & 534.80 & -1.07 & 33.74 & 1.81 & - & - & \cellcolor{green}15.85 & -0.59 \\ 
135 & Mn$_{4}$Si$_{4}$ & 198 & 147.21 & -0.76 & 138.30 & -0.73 & 150~\cite{liu2018giant} & -0.80 at ... K~\cite{yang2020giant} & 1.06 & 1.05 \\ 
\rowcolor{Gray}
136 & Fe$_{4}$Ge$_{4}$ & 198 & 121.17 & 0.00 & 147.02 & 0.01 & - & - & 0.82 & 0.10 \\ 
137 & Fe$_{4}$Sb$_{4}$Pt$_{4}$ & 198 & 208.60 & -1.72 & 177.74 & -1.63 & - & - & 1.17 & 1.05 \\ 
\rowcolor{Gray}
138 & Fe$_{4}$Si$_{4}$ & 198 & 75.57 & 0.12 & 14.09 & 0.16 & - & - & \cellcolor{green}5.36 & 0.70 \\ 
139 & Na$_{1}$Fe$_{4}$Sb$_{12}$ & 204 & 617.22 & \cellcolor{cyan}-3.18 & 850.48 & -2.78 & - & - & 0.73 & 1.14 \\ 
\rowcolor{Gray}
140 & S$_{8}$Co$_{4}$ & 205 & -365.66 & 2.15 & -351.35 & 2.19 & - & - & 1.04 & 0.98 \\ 
141 & Co$_{4}$Se$_{8}$ & 205 & 510.92 & -1.22 & 196.23 & -1.59 & - & - & 2.60 & 0.77 \\ 
\rowcolor{Gray}
142 & Se$_{8}$Ni$_{4}$ & 205 & 1.42 & -0.08 & 10.07 & 0.07 & - & - & \cellcolor{green}0.14 & -1.23 \\ 
143 & Pd$_{8}$Mo$_{12}$N$_{4}$ & 213 & -3.28 & -0.07 & 146.90 & 0.14 & - & - & \cellcolor{green}-0.02 & -0.49 \\ 
\rowcolor{Gray}
144 & Mn$_{1}$Ni$_{1}$Sb$_{1}$ & 216 & 164.64 & 0.09 & 164.65 & 0.09 & - & 0.23 at 500K~\cite{sakai2020iron} & 1.00 & 1.06 \\ 
145 & Ti$_{1}$Co$_{1}$Sn$_{1}$ & 216 & 150.77 & 0.11 & 149.37 & 0.18 & - & 0.23 at 500K~\cite{sakai2020iron} & 1.01 & 0.65 \\ 
\rowcolor{Gray}
146 & Mn$_{1}$Co$_{1}$Sb$_{1}$ & 216 & 2.36 & -0.39 & 20.23 & -0.30 & - & 0.51 at 500K~\cite{sakai2020iron} & \cellcolor{green}0.12 & 1.31 \\ 
147 & Mn$_{1}$Sn$_{1}$Pt$_{1}$ & 216 & 138.68 & 0.85 & -42.11 & 1.06 & - & 1.70 at 500K~\cite{sakai2020iron} & -3.29 & 0.80 \\ 
\rowcolor{Gray}
148 & Mn$_{1}$Sb$_{1}$Au$_{1}$ & 216 & 397.47 & 0.45 & 424.26 & 0.09 & - & 0.99 at 500K~\cite{sakai2020iron} & 0.94 & \cellcolor{green}5.30 \\ 
149 & Mn$_{1}$Sb$_{1}$Pt$_{1}$ & 216 & 81.37 & 0.50 & 143.12 & 0.39 & - & 0.68 at 500K~\cite{sakai2020iron} & 0.57 & 1.29 \\ 
\rowcolor{Gray}
150 & V$_{1}$Co$_{1}$Sb$_{1}$ & 216 & 103.27 & 0.40 & 59.07 & 0.51 & - & 0.47 at 500K~\cite{sakai2020iron} & 1.75 & 0.77 \\ 
151 & Mn$_{1}$Sn$_{1}$Au$_{1}$ & 216 & -58.80 & 0.18 & -50.28 & 0.60 & - & 0.39 at 500K~\cite{sakai2020iron} & 1.17 & 0.30 \\ 
\rowcolor{Gray}
152 & Bi$_{8}$Mn$_{10}$Ni$_{4}$ & 216 & 321.82 & -1.63 & 329.89 & -1.09 & - & - & 0.98 & 1.49 \\ 
153 & Al$_{4}$Cr$_{4}$Fe$_{4}$Co$_{4}$ & 216 & -250.90 & -0.03 & -129.23 & 0.61 & - & - & 1.94 & \cellcolor{green}-0.04 \\ 
\rowcolor{Gray}
154 & Cr$_{4}$Cu$_{1}$Se$_{8}$In$_{1}$ & 216 & 36.00 & -1.14 & 57.99 & -1.10 & - & - & 0.62 & 1.03 \\ 
155 & Rh$_{4}$Mn$_{4}$Sn$_{4}$ & 216 & 49.09 & 0.81 & 28.53 & 0.81 & - & - & 1.72 & 0.99 \\ 
\rowcolor{Gray}
156 & Mn$_{1}$Fe$_{1}$Co$_{1}$Ge$_{1}$ & 216 & 23.56 & 1.27 & 21.73 & 1.18 & - & - & 1.08 & 1.07 \\ 
157 & Mn$_{2}$Co$_{1}$Sb$_{1}$ & 216 & 29.20 & -0.11 & 28.58 & -0.07 & - & 0.67 at 500K~\cite{sakai2020iron} & 1.02 & 1.73 \\ 
\rowcolor{Gray}
158 & Mn$_{2}$Fe$_{1}$C$_{6}$N$_{6}$ & 216 & 0.00 & 0.19 & 0.00 & -0.40 & - & - & 0.40 & -0.49 \\ 
159 & Zr$_{1}$In$_{1}$Ni$_{4}$ & 216 & 335.04 & -0.02 & 363.00 & -0.27 & - & - & 0.92 & \cellcolor{green}0.09 \\ 
\rowcolor{Gray}
160 & Ni$_{5}$Zr$_{1}$ & 216 & -220.27 & 1.50 & -202.99 & 1.76 & - & 2.43 at 500K~\cite{sakai2020iron} & 1.09 & 0.85 \\ 
161 & As$_{12}$ & 220 & 891.79 & 1.24 & 536.22 & 2.98 & - & - & 1.66 & 0.42 \\ 
\rowcolor{Gray}
162 & P$_{12}$ & 220 & 0.00 & -1.08 & 0.00 & -1.08 & - & - & 2.69 & 1.00 \\ 
163 & Al$_{1}$Ni$_{3}$ & 221 & 206.19 & -2.74 & 208.74 & -2.15 & - & 3.35 at 500K~\cite{sakai2020iron} & 0.99 & 1.28 \\ 
\rowcolor{Gray}
164 & Fe$_{1}$Pt$_{3}$ & 221 & -318.04 & -1.83 & -318.40 & -1.73 & - & - & 1.00 & 1.06 \\ 
165 & Cr$_{1}$Pt$_{3}$ & 221 & \cellcolor{cyan}2060.08 & -0.10 & \cellcolor{cyan}2040.21 & -0.24 & 2040~\cite{markou2020hard} & 2.75 at 500K~\cite{sakai2020iron} & 1.01 & 0.42 \\ 
\rowcolor{Gray}
166 & Al$_{1}$Fe$_{1}$ & 221 & -302.83 & -1.59 & -310.86 & -1.62 & - & 3.34 at 500K~\cite{sakai2020iron} & 0.97 & 0.98 \\ 
167 & Fe$_{1}$Ni$_{3}$ & 221 & -127.38 & 0.02 & -80.97 & -0.01 & - & 2.84 at 500K~\cite{sakai2020iron} & 1.57 & -1.17 \\ 
\rowcolor{Gray}
168 & Fe$_{1}$Rh$_{1}$ & 221 & 809.19 & -0.13 & 794.39 & -0.12 & - & 1.98 at 500K~\cite{sakai2020iron} & 1.02 & 1.05 \\ 
169 & Ga$_{1}$Co$_{1}$ & 221 & -9.13 & 0.08 & -16.84 & 0.08 & - & - & 0.54 & 0.95 \\ 
\rowcolor{Gray}
170 & Fe$_{1}$Pd$_{3}$ & 221 & -201.61 & 1.56 & -217.08 & 1.61 & - & 2.20 at 500K~\cite{sakai2020iron} & 0.93 & 0.97 \\ 
171 & Mn$_{1}$Pt$_{3}$ & 221 & \cellcolor{cyan}1342.88 & \cellcolor{cyan}-4.06 & \cellcolor{cyan}1471.55 & \cellcolor{cyan}-4.58 & 1400~\cite{markou2020hard} & 4.55 at 500K~\cite{sakai2020iron} & 0.91 & 0.89 \\ 
\rowcolor{Gray}
172 & Mn$_{1}$Zn$_{1}$ & 221 & 339.81 & -1.64 & \cellcolor{cyan}1059.69 & -0.78 & - & 2.40 at 500K~\cite{sakai2020iron} & 0.32 & 2.10 \\ 
173 & Fe$_{4}$N$_{1}$ & 221 & 631.04 & 0.45 & 557.85 & -0.60 & - & 2.44 at 500K~\cite{sakai2020iron} & 1.13 & -0.75 \\ 
\rowcolor{Gray}
174 & Mn$_{1}$Ni$_{3}$ & 221 & 45.87 & 0.94 & 43.37 & 0.76 & - & - & 1.06 & 1.23 \\ 
175 & Mn$_{4}$N$_{1}$ & 221 & 119.34 & -0.04 & 884.35 & 1.93 & - & 0.83 at 500K~\cite{sakai2020iron} & \cellcolor{green}0.13 & \cellcolor{green}-0.02 \\ 
\rowcolor{Gray}
176 & V$_{1}$Fe$_{1}$ & 221 & 61.81 & -2.93 & 59.98 & -1.80 & - & 2.56 at 500K~\cite{sakai2020iron} & 1.03 & 1.62 \\ 
177 & Co$_{1}$Pt$_{3}$ & 221 & -978.05 & -0.50 & -933.11 & -0.58 & - & 1.04 at 500K~\cite{sakai2020iron} & 1.05 & 0.87 \\ 
\rowcolor{Gray}
178 & Mn$_{3}$Ge$_{1}$ & 221 & -130.51 & \cellcolor{cyan}4.78 & -57.15 & \cellcolor{cyan}4.93 & - & 1.38 at 500K~\cite{sakai2020iron} & 2.28 & 0.97 \\ 
179 & C$_{1}$Mn$_{3}$In$_{1}$ & 221 & -52.54 & -0.98 & -157.79 & 0.63 & - & - & 0.33 & -1.57 \\ 
\rowcolor{Gray}
180 & Ni$_{3}$Pt$_{1}$ & 221 & \cellcolor{cyan}-2044.52 & \cellcolor{cyan}-7.24 & \cellcolor{cyan}-2051.84 & \cellcolor{cyan}-7.29 & - & - & 1.00 & 0.99 \\ 
181 & Ti$_{1}$Co$_{3}$ & 221 & -125.38 & -1.46 & -139.69 & -1.61 & - & - & 0.90 & 0.90 \\ 
\rowcolor{Gray}
182 & B$_{6}$Ba$_{1}$ & 221 & -0.80 & 0.01 & 51.03 & -0.12 & - & - & \cellcolor{green}-0.02 & \cellcolor{green}-0.07 \\ 
183 & C$_{1}$Al$_{1}$Mn$_{3}$ & 221 & 181.58 & -0.48 & 180.17 & -0.50 & - & 0.93 at 500K~\cite{sakai2020iron} & 1.01 & 0.96 \\ 
\rowcolor{Gray}
184 & Ni$_{3}$Al$_{1}$C$_{1}$ & 221 & -4.68 & 0.02 & -3.32 & 0.04 & - & - & 1.41 & 0.56 \\ 
185 & Mn$_{1}$Ni$_{2}$Ga$_{1}$ & 225 & 9.84 & 0.41 & -41.73 & 0.46 & 31~\cite{noky2020giant} & -0.50 at 300K~\cite{noky2020giant} & \cellcolor{green}-0.24 & 0.89 \\ 
\rowcolor{Gray}
186 & Mn$_{1}$Ni$_{2}$Sn$_{1}$ & 225 & -83.45 & -0.83 & -60.95 & -0.80 & -9~\cite{noky2020giant} & 0.66 at 300K~\cite{noky2020giant} & 1.37 & 1.03 \\ 
187 & Ni$_{1}$ & 225 & \cellcolor{cyan}-2038.91 & 0.99 & \cellcolor{cyan}-2444.07 & 0.76 & -2500~\cite{weischenberg2011ab} & 0.22 at 300K~\cite{hasegawa2015material} & 0.83 & 1.29 \\ 
\rowcolor{Gray}
188 & Mn$_{1}$Ni$_{2}$In$_{1}$ & 225 & -50.86 & 0.20 & -60.36 & -0.03 & -99~\cite{noky2020giant} & -0.49 at 300K~\cite{noky2020giant} & 0.84 & \cellcolor{green}-7.38 \\ 
189 & Mn$_{1}$Ni$_{2}$Sb$_{1}$ & 225 & -60.65 & 0.06 & -86.11 & 0.04 & - & 0.26 at 500K~\cite{sakai2020iron} & 0.70 & 1.52 \\ 
\rowcolor{Gray}
190 & Mn$_{1}$Pd$_{2}$Sn$_{1}$ & 225 & 110.43 & 0.86 & 122.81 & 0.89 & - & 1.21 at 500K~\cite{sakai2020iron} & 0.90 & 0.97 \\ 
191 & Ti$_{1}$Co$_{2}$Sn$_{1}$ & 225 & 130.36 & 0.04 & 134.33 & 0.03 & -114~\cite{noky2020giant} & -0.02 at 300K~\cite{noky2020giant} & 0.97 & 1.34 \\ 
\rowcolor{Gray}
192 & Al$_{1}$Fe$_{3}$ & 225 & 216.26 & -1.54 & 218.06 & -1.65 & -265~\cite{noky2020giant} & 1.33 at 300K~\cite{noky2020giant} & 0.99 & 0.93 \\ 
193 & Mn$_{1}$Co$_{2}$Sn$_{1}$ & 225 & 180.14 & -0.00 & 200.08 & -0.00 & 118~\cite{kubler2012berry} & 0.11 at 300K~\cite{noky2020giant} & 0.90 & 0.62 \\ 
\rowcolor{Gray}
194 & Al$_{1}$Ti$_{1}$Co$_{2}$ & 225 & 112.53 & -0.31 & 77.67 & -0.56 & -84~\cite{noky2020giant} & 0.31 at 300K~\cite{noky2020giant} & 1.45 & 0.56 \\ 
195 & Ga$_{1}$Fe$_{3}$ & 225 & 743.55 & -1.71 & 797.26 & -1.65 & - & 5.22 at 500K~\cite{sakai2020iron} & 0.93 & 1.04 \\ 
\rowcolor{Gray}
196 & Fe$_{3}$Si$_{1}$ & 225 & 341.40 & -0.28 & 331.33 & -0.11 & -330~\cite{noky2020giant} & 0.59 at 300K~\cite{noky2020giant} & 1.03 & 2.48 \\ 
197 & Mn$_{1}$Rh$_{2}$Sn$_{1}$ & 225 & 293.61 & 0.46 & 398.36 & 0.48 & -270~\cite{noky2020giant} & 0.31 at 300K~\cite{noky2020giant} & 0.74 & 0.95 \\ 
\rowcolor{Gray}
198 & Mn$_{1}$Pd$_{2}$Sb$_{1}$ & 225 & 376.96 & -0.76 & 314.12 & -0.90 & - & 1.38 at 500K~\cite{sakai2020iron} & 1.20 & 0.84 \\ 
199 & Mn$_{1}$Al$_{1}$Cu$_{2}$ & 225 & 185.04 & -0.18 & 168.81 & -0.10 & - & 0.23 at 500K~\cite{sakai2020iron} & 1.10 & 1.79 \\ 
\rowcolor{Gray}
200 & Fe$_{1}$Co$_{2}$Si$_{1}$ & 225 & -138.85 & \cellcolor{cyan}-4.85 & -509.80 & \cellcolor{cyan}-6.69 & -275~\cite{noky2020giant} & 2.57 at 300K~\cite{noky2020giant} & 0.27 & 0.72 \\ 
201 & Al$_{1}$V$_{1}$Fe$_{2}$ & 225 & 8.61 & 0.06 & 8.48 & 0.08 & - & - & 1.02 & 0.73 \\ 
\rowcolor{Gray}
202 & Fe$_{23}$Y$_{6}$ & 225 & 231.13 & 1.68 & 272.00 & 1.03 & - & - & 0.85 & 1.62 \\ 
203 & In$_{2}$Au$_{1}$ & 225 & 3.80 & -0.01 & 84.77 & 0.00 & - & - & \cellcolor{green}0.04 & -4.93 \\ 
\rowcolor{Gray}
204 & Mn$_{6}$Cu$_{8}$Bi$_{8}$ & 225 & 714.73 & -0.10 & 718.54 & 1.02 & - & - & 0.99 & \cellcolor{green}-0.10 \\ 
205 & V$_{1}$Co$_{2}$Sn$_{1}$ & 225 & 108.60 & -1.23 & -79.60 & -1.42 & 164~\cite{noky2020giant} & 0.63 at 300K~\cite{noky2020giant} & -1.36 & 0.86 \\ 
\rowcolor{Gray}
206 & Al$_{1}$Co$_{2}$Zr$_{1}$ & 225 & 126.80 & -0.34 & 157.87 & -0.44 & - & 0.73 at 500K~\cite{sakai2020iron} & 0.80 & 0.77 \\ 
207 & Zr$_{1}$Co$_{2}$Sn$_{1}$ & 225 & 75.28 & 0.01 & 83.65 & 0.01 & - & 0.05 at 500K~\cite{sakai2020iron} & 0.90 & 1.16 \\ 
\rowcolor{Gray}
208 & Al$_{1}$Mn$_{1}$Au$_{2}$ & 225 & 84.91 & -0.16 & 80.77 & -0.53 & - & 0.83 at 500K~\cite{sakai2020iron} & 1.05 & 0.30 \\ 
209 & Al$_{1}$Cr$_{1}$Co$_{2}$ & 225 & 815.94 & 1.17 & 657.72 & 1.34 & 438~\cite{kubler2012berry} & 3.23 at 300K~\cite{noky2020giant} & 1.24 & 0.87 \\ 
\rowcolor{Gray}
210 & Fe$_{8}$Cr$_{4}$Si$_{4}$ & 225 & 720.50 & -2.33 & 892.13 & -2.83 & - & - & 0.81 & 0.82 \\ 
211 & Mn$_{4}$Sn$_{4}$Cu$_{8}$ & 225 & 198.14 & 0.40 & 190.94 & 0.31 & - & - & 1.04 & 1.26 \\ 
\rowcolor{Gray}
212 & Ga$_{1}$Fe$_{2}$Ni$_{1}$ & 225 & \cellcolor{cyan}1239.52 & -1.49 & \cellcolor{cyan}1487.77 & -2.33 & -69~\cite{noky2020giant} & 1.27 at 300K~\cite{noky2020giant} & 0.83 & 0.64 \\ 
213 & Fe$_{4}$Ru$_{8}$Si$_{4}$ & 225 & \cellcolor{cyan}1044.72 & -1.12 & \cellcolor{cyan}1037.94 & -1.67 & -1054~\cite{noky2020giant} & 0.38 at 300K~\cite{noky2020giant} & 1.01 & 0.67 \\ 
\rowcolor{Gray}
214 & Co$_{2}$Sn$_{1}$Hf$_{1}$ & 225 & 43.87 & 0.58 & 69.77 & 0.43 & - & 0.19 at 500K~\cite{sakai2020iron} & 0.63 & 1.35 \\ 
215 & Mn$_{1}$Co$_{2}$Si$_{1}$ & 225 & 194.27 & -0.50 & 166.22 & -0.55 & -187~\cite{noky2020giant} & 0.66 at 300K~\cite{noky2020giant} & 1.17 & 0.90 \\ 
\rowcolor{Gray}
216 & Al$_{1}$Mn$_{1}$Ni$_{2}$ & 225 & -162.87 & 0.91 & -128.55 & 0.80 & 223~\cite{noky2020giant} & -0.96 at 300K~\cite{noky2020giant} & 1.27 & 1.13 \\ 
217 & Mn$_{1}$Ni$_{2}$Ge$_{1}$ & 225 & 110.38 & -1.05 & 104.19 & -1.07 & -155~\cite{noky2020giant} & 1.12 at 300K~\cite{noky2020giant} & 1.06 & 0.99 \\ 
\rowcolor{Gray}
218 & Mn$_{1}$Pd$_{2}$Ge$_{1}$ & 225 & 289.20 & 0.35 & 294.26 & 0.45 & - & 1.01 at 500K~\cite{sakai2020iron} & 0.98 & 0.76 \\ 
219 & Mn$_{1}$Rh$_{2}$Pb$_{1}$ & 225 & -119.31 & -0.93 & -297.19 & -1.09 & - & 1.67 at 500K~\cite{sakai2020iron} & 0.40 & 0.86 \\ 
\rowcolor{Gray}
220 & Ni$_{1}$Ge$_{1}$Rh$_{2}$ & 225 & -0.00 & 0.00 & -0.07 & 0.00 & - & - & 0.01 & 0.14 \\ 
221 & Ti$_{1}$Co$_{2}$Ga$_{1}$ & 225 & 146.02 & -0.68 & 149.46 & -0.83 & -46~\cite{noky2020giant} & 0.55 at 300K~\cite{noky2020giant} & 0.98 & 0.81 \\ 
\rowcolor{Gray}
222 & Al$_{1}$Ti$_{1}$Fe$_{2}$ & 225 & 72.93 & -0.28 & 69.22 & -0.24 & -38~\cite{noky2020giant} & 0.03 at 300K~\cite{noky2020giant} & 1.05 & 1.16 \\ 
223 & Al$_{1}$V$_{1}$Mn$_{2}$ & 225 & 59.25 & -0.47 & 70.56 & -0.40 & -51~\cite{noky2020giant} & 0.36 at 300K~\cite{noky2020giant} & 0.84 & 1.17 \\ 
\rowcolor{Gray}
224 & Cr$_{1}$Ga$_{1}$Co$_{2}$ & 225 & 618.69 & 1.64 & 654.20 & 1.81 & -645~\cite{noky2020giant} & 2.02 at 300K~\cite{noky2020giant} & 0.95 & 0.91 \\ 
225 & Al$_{1}$Cr$_{1}$Fe$_{2}$ & 225 & -120.08 & -0.88 & 103.73 & 0.11 & -155~\cite{noky2020giant} & -0.18 at 300K~\cite{noky2020giant} & -1.16 & \cellcolor{green}-8.21 \\ 
\rowcolor{Gray}
226 & Cr$_{1}$Fe$_{2}$Ga$_{1}$ & 225 & -166.65 & 2.76 & -138.68 & 1.03 & -190~\cite{noky2020giant} & -0.68 at 300K~\cite{noky2020giant} & 1.20 & 2.67 \\ 
227 & N$_{2}$Cr$_{2}$ & 225 & -171.32 & -1.69 & -123.82 & -0.71 & - & - & 1.38 & 2.39 \\ 
\rowcolor{Gray}
228 & Al$_{1}$Cr$_{1}$Ni$_{2}$ & 225 & -498.01 & -1.06 & -512.11 & -1.00 & 286~\cite{noky2020giant} & 1.83 at 300K~\cite{noky2020giant} & 0.97 & 1.06 \\ 
229 & Mn$_{1}$Cu$_{2}$In$_{1}$ & 225 & 57.88 & 0.34 & 41.52 & 0.49 & - & 0.74 at 500K~\cite{sakai2020iron} & 1.39 & 0.69 \\ 
\rowcolor{Gray}
230 & Fe$_{1}$Sn$_{1}$Ru$_{2}$ & 225 & 201.04 & -1.22 & 204.57 & -0.99 & -232~\cite{noky2020giant} & -1.66 at 300K~\cite{noky2020giant} & 0.98 & 1.23 \\ 
231 & Ga$_{1}$Fe$_{2}$Co$_{1}$ & 225 & -227.40 & 2.19 & -208.68 & \cellcolor{cyan}3.28 & -54~\cite{noky2020giant} & 1.05 at 300K~\cite{noky2020giant} & 1.09 & 0.67 \\ 
\rowcolor{Gray}
232 & Al$_{1}$Co$_{2}$Hf$_{1}$ & 225 & 216.34 & -0.24 & 250.96 & -0.17 & - & - & 0.86 & 1.42 \\ 
233 & Mn$_{1}$Ga$_{2}$Co$_{1}$ & 225 & 32.25 & -1.33 & 120.81 & -0.72 & - & - & 0.27 & 1.84 \\ 
\rowcolor{Gray}
234 & Mn$_{1}$Co$_{2}$Ge$_{1}$ & 225 & 307.22 & -0.11 & 306.85 & -0.12 & -253~\cite{noky2020giant} & 0.27 at 300K~\cite{noky2020giant} & 1.00 & 0.93 \\ 
235 & Mn$_{1}$Co$_{2}$Sb$_{1}$ & 225 & -7.43 & -0.97 & -11.80 & -0.98 & -4~\cite{noky2020giant} & 0.55 at 300K~\cite{noky2020giant} & 0.63 & 0.99 \\ 
\rowcolor{Gray}
236 & Al$_{1}$Mn$_{1}$Fe$_{2}$ & 225 & 743.00 & 1.74 & 731.70 & 1.52 & - & - & 1.02 & 1.14 \\ 
237 & Mn$_{1}$Fe$_{2}$Si$_{1}$ & 225 & 118.95 & 0.68 & 163.44 & 0.75 & -118~\cite{noky2020giant} & -0.82 at 300K~\cite{noky2020giant} & 0.73 & 0.91 \\ 
\rowcolor{Gray}
238 & Al$_{1}$Co$_{2}$Nb$_{1}$ & 225 & 28.90 & 0.17 & 31.29 & 0.23 & - & 0.46 at 500K~\cite{sakai2020iron} & 0.92 & 0.74 \\ 
239 & Ni$_{1}$Rh$_{2}$Sn$_{1}$ & 225 & 20.26 & \cellcolor{cyan}-9.11 & 105.76 & \cellcolor{cyan}-9.06 & 360~\cite{noky2020giant} & 8.14 at 300K~\cite{noky2020giant} & \cellcolor{green}0.19 & 1.01 \\ 
\rowcolor{Gray}
240 & Sc$_{1}$Co$_{2}$Sn$_{1}$ & 225 & 129.93 & -0.41 & 137.74 & -0.38 & -63~\cite{noky2020giant} & 0.2 at 300K~\cite{noky2020giant} & 0.94 & 1.08 \\ 
241 & Sr$_{8}$Ru$_{4}$ & 225 & -88.51 & 1.19 & -106.15 & 1.00 & - & - & 0.83 & 1.18 \\ 
\rowcolor{Gray}
242 & Al$_{1}$Co$_{2}$Ta$_{1}$ & 225 & -76.07 & 0.75 & -72.56 & 0.70 & - & - & 1.05 & 1.06 \\ 
243 & Ti$_{1}$Co$_{2}$Si$_{1}$ & 225 & 151.41 & 0.58 & 113.08 & 0.36 & -118~\cite{noky2020giant} & -0.22 at 300K~\cite{noky2020giant} & 1.34 & 1.63 \\ 
\rowcolor{Gray}
244 & Al$_{1}$V$_{1}$Co$_{2}$ & 225 & 151.02 & 0.15 & 167.09 & 0.10 & -166~\cite{noky2020giant} & -0.28 at 300K~\cite{noky2020giant} & 0.90 & 1.44 \\ 
245 & V$_{1}$Co$_{2}$Ga$_{1}$ & 225 & 142.59 & 0.06 & 140.34 & 0.03 & 137~\cite{manna2018colossal} & -0.16 at 300K~\cite{noky2020giant} & 1.02 & 2.39 \\ 
\rowcolor{Gray}
246 & Si$_{1}$V$_{1}$Co$_{2}$ & 225 & 309.01 & -1.06 & 284.71 & -1.40 & - & 1.06 at 500K~\cite{sakai2020iron} & 1.09 & 0.76 \\ 
247 & Si$_{1}$V$_{1}$Fe$_{2}$ & 225 & 39.61 & 1.00 & 35.00 & 0.78 & -147~\cite{noky2020giant} & -1.54 at 300K~\cite{noky2020giant} & 1.13 & 1.28 \\ 
\rowcolor{Gray}
248 & V$_{1}$Fe$_{2}$Sn$_{1}$ & 225 & 330.15 & 0.34 & 307.39 & 1.03 & 78~\cite{noky2020giant} & -0.23 at 300K~\cite{noky2020giant} & 1.07 & 0.33 \\ 
249 & Mn$_{2}$V$_{1}$Ga$_{1}$ & 225 & 63.89 & 1.02 & 109.90 & 1.06 & 29~\cite{noky2020giant} & -0.64 at 300K~\cite{noky2020giant} & 0.58 & 0.96 \\ 
\rowcolor{Gray}
250 & Mn$_{2}$Sn$_{1}$W$_{1}$ & 225 & 49.45 & \cellcolor{cyan}-4.99 & 708.77 & -2.23 & - & - & \cellcolor{green}0.07 & 2.24 \\ 
251 & Zn$_{4}$Zr$_{2}$ & 227 & -165.25 & -0.47 & -181.43 & -0.33 & - & - & 0.91 & 1.40 \\ 
\rowcolor{Gray}
252 & Y$_{2}$Fe$_{4}$ & 227 & 4.82 & -0.97 & 41.57 & -0.75 & - & - & \cellcolor{green}0.12 & 1.29 \\ 
253 & Cr$_{4}$Se$_{8}$Cd$_{2}$ & 227 & 0.00 & -0.83 & 0.00 & -0.79 & - & - & 70.09 & 1.05 \\ 
\rowcolor{Gray}
254 & S$_{8}$Cr$_{4}$Cd$_{2}$ & 227 & -0.83 & 0.10 & -0.63 & 0.17 & - & - & 1.32 & 0.59 \\ 
255 & Fe$_{4}$Zr$_{2}$ & 227 & 215.31 & 0.90 & 237.03 & 1.10 & - & - & 0.91 & 0.82 \\ 
\rowcolor{Gray}
256 & Cr$_{4}$Cu$_{2}$Se$_{8}$ & 227 & 28.35 & -0.41 & 13.44 & -0.46 & - & - & 2.11 & 0.89 \\ 
257 & S$_{8}$Cr$_{4}$Fe$_{2}$ & 227 & -280.51 & 2.08 & -51.36 & 2.40 & - & - & \cellcolor{green}5.46 & 0.87 \\ 
\rowcolor{Gray}
258 & Cr$_{4}$Se$_{8}$Hg$_{2}$ & 227 & 137.25 & -0.91 & 30.13 & -0.76 & 35~\cite{yang2019unconventional} & - & \cellcolor{green}4.55 & 1.20 \\ 
259 & Cr$_{4}$Cu$_{2}$Te$_{8}$ & 227 & -47.44 & -0.35 & -145.88 & -0.46 & - & - & 0.33 & 0.75 \\ 
\rowcolor{Gray}
260 & S$_{8}$Cr$_{4}$Cu$_{2}$ & 227 & -24.72 & -0.98 & -40.91 & -1.07 & - & - & 0.60 & 0.91 \\ 
261 & Co$_{4}$Zr$_{2}$ & 227 & -147.93 & 1.10 & -163.83 & 1.42 & - & - & 0.90 & 0.78 \\ 
\rowcolor{Gray}
262 & S$_{8}$Cr$_{4}$Fe$_{1}$Cu$_{1}$ & 227 & -420.86 & 0.11 & -140.60 & 0.27 & - & - & 2.99 & 0.39 \\ 
263 & Fe$_{4}$F$_{12}$ & 227 & 0.00 & 0.05 & 0.03 & -0.00 & - & - & 0.00 & -20.80 \\ 
\rowcolor{Gray}
264 & Fe$_{6}$S$_{8}$ & 227 & -353.69 & 2.47 & 837.63 & -2.56 & - & - & -0.42 & -0.97 \\ 
265 & Y$_{2}$Co$_{4}$ & 227 & 662.69 & -1.05 & 855.03 & -1.32 & - & - & 0.78 & 0.80 \\ 
\rowcolor{Gray}
266 & Fe$_{1}$ & 229 & 862.83 & -0.69 & 833.57 & -0.92 & 767~\cite{weischenberg2011ab} & 4.08 at 500K~\cite{sakai2020iron} & 1.04 & 0.76 \\ 
\hline
\caption{AHC at T=0K and ANC at T=300K for the two magnetization directions ($\sigma_x$, $\alpha_x$ for $\mathbf{M}//\mathbf{x}$ and $\sigma_z$, $\alpha_z$ for $\mathbf{M}//\mathbf{z}$), comparison with the literature and the anisotropy of each ferromagnetic compound. The choice of different unit cell between our results and the reported cases is possible. Values highlighted in blue signify the most interesting cases while the ones in green exhibit the largest anisotropy in either direction.}
\label{tab1}
\end{longtable}
\end{tiny}

\newpage

\begin{tiny}
\begin{longtable}{|c|c|c|}
\hline
\multicolumn{3}{|c|}{Magnetic Laue group: $4/mm^{`}m^{`}$}\\
\hline
Symmetry & General & Berry Curvature \\
                  & Position & \\
\hline
 &  & $\Omega_{x}\left(k_x,k_y,k_z\right) = \Omega_{x}\left(k_x,k_y,k_z\right)$ \\
$1$  & $x,y,z$ & $\Omega_{y}\left(k_x,k_y,k_z\right) = \Omega_{y}\left(k_x,k_y,k_z\right)$ \\
 &  & $\Omega_{z}\left(k_x,k_y,k_z\right) = \Omega_{z}\left(k_x,k_y,k_z\right)$ \\
\hline
 &  & $\Omega_{x}\left(-k_y,k_x,k_z\right) = \Omega_{y}\left(k_x,k_y,k_z\right)$ \\
$4_{z}^{+}$ & $-y,x,z$ & $\Omega_{y}\left(-k_y,k_x,k_z\right) = -\Omega_{x}\left(k_x,k_y,k_z\right)$ \\
 &  & $\Omega_{z}\left(-k_y,k_x,k_z\right) = \Omega_{z}\left(k_x,k_y,k_z\right)$ \\
\hline
 &  & $\Omega_{x}\left(k_y,-k_x,k_z\right) = -\Omega_{y}\left(k_x,k_y,k_z\right)$ \\
$4_{z}^{-}$ & $y,-x,z$ & $\Omega_{y}\left(k_y,-k_x,k_z\right) = \Omega_{x}\left(k_x,k_y,k_z\right)$ \\
 &  & $\Omega_{z}\left(k_y,-k_x,k_z\right) = \Omega_{z}\left(k_x,k_y,k_z\right)$ \\
\hline
 &  & $\Omega_{x}\left(-k_x,k_y,k_z\right) = -\Omega_{x}\left(k_x,k_y,k_z\right)$ \\
$2_x'$ & $x,-y,-z$ & $\Omega_{y}\left(-k_x,k_y,k_z\right) = \Omega_{y}\left(k_x,k_y,k_z\right)$ \\
 &  & $\Omega_{z}\left(-k_x,k_y,k_z\right) = \Omega_{z}\left(k_x,k_y,k_z\right)$ \\
\hline
 &  & $\Omega_{x}\left(k_x,-k_y,k_z\right) = \Omega_{x}\left(k_x,k_y,k_z\right)$ \\
$2_y'$ & $-x,y,-z$ & $\Omega_{y}\left(k_x,-k_y,k_z\right) = -\Omega_{y}\left(k_x,k_y,k_z\right)$ \\
 &  & $\Omega_{z}\left(k_x,-k_y,k_z\right) = \Omega_{z}\left(k_x,k_y,k_z\right)$ \\
\hline
 &  & $\Omega_{x}\left(-k_x,-k_y,k_z\right) = -\Omega_{x}\left(k_x,k_y,k_z\right)$ \\
 $2_{z}$ & $-x,-y,z$ & $\Omega_{y}\left(-k_x,-k_y,k_z\right) = -\Omega_{y}\left(k_x,k_y,k_z\right)$ \\
 &  & $\Omega_{z}\left(-k_x,-k_y,k_z\right) = \Omega_{z}\left(k_x,k_y,k_z\right)$ \\
\hline
 &  & $\Omega_{x}\left(-k_y,-k_x,k_z\right) = -\Omega_{y}\left(k_x,k_y,k_z\right)$ \\ 
 $2_{xy}'$ & $y,x,-z$ & $\Omega_{y}\left(-k_y,-k_x,k_z\right) = -\Omega_{x}\left(k_x,k_y,k_z\right)$ \\
 &  & $\Omega_{z}\left(-k_y,-k_x,k_z\right) = \Omega_{z}\left(k_x,k_y,k_z\right)$ \\
\hline
 &  & $\Omega_{x}\left(k_y,k_x,k_z\right) = \Omega_{y}\left(k_x,k_y,k_z\right)$ \\ 
$2_{x-y}'$ & $-y,-x,-z$ & $\Omega_{y}\left(k_y,k_x,k_z\right) = \Omega_{x}\left(k_x,k_y,k_z\right)$ \\
 &  & $\Omega_{z}\left(k_y,k_x,k_z\right) = \Omega_{z}\left(k_x,k_y,k_z\right)$ \\
\hline
 &  & $\Omega_{x}\left(-k_x,-k_y,-k_z\right) = \Omega_{x}\left(k_x,k_y,k_z\right)$ \\ 
 $-1$ & $-x,-y,-z$ & $\Omega_{y}\left(-k_x,-k_y,-k_z\right) = \Omega_{y}\left(k_x,k_y,k_z\right)$ \\
 &  & $\Omega_{z}\left(-k_x,-k_y,-k_z\right) = \Omega_{z}\left(k_x,k_y,k_z\right)$ \\
\hline
 &  & $\Omega_{x}\left(k_y,-k_x,-k_z\right) = \Omega_{y}\left(k_x,k_y,k_z\right)$ \\ 
$-4_{z}^{+}$ & $y,-x,-z$ & $\Omega_{y}\left(k_y,-k_x,-k_z\right) = -\Omega_{x}\left(k_x,k_y,k_z\right)$ \\
 &  & $\Omega_{z}\left(k_y,-k_x,-k_z\right) = \Omega_{z}\left(k_x,k_y,k_z\right)$ \\
\hline
 &  & $\Omega_{x}\left(-k_y,k_x,-k_z\right) = -\Omega_{y}\left(k_x,k_y,k_z\right)$ \\ 
$-4_{z}^{-}$ & $-y,x,-z$ & $\Omega_{y}\left(-k_y,k_x,-k_z\right) = \Omega_{x}\left(k_x,k_y,k_z\right)$ \\
 &  & $\Omega_{z}\left(-k_y,k_x,-k_z\right) = \Omega_{z}\left(k_x,k_y,k_z\right)$ \\
\hline
 &  & $\Omega_{x}\left(k_x,-k_y,-k_z\right) = -\Omega_{x}\left(k_x,k_y,k_z\right)$ \\ 
 $m_{x}'$ & $-x,y,z$ & $\Omega_{y}\left(k_x,-k_y,-k_z\right) = \Omega_{y}\left(k_x,k_y,k_z\right)$ \\
 &  & $\Omega_{z}\left(k_x,-k_y,-k_z\right) = \Omega_{z}\left(k_x,k_y,k_z\right)$ \\
\hline
 &  & $\Omega_{x}\left(-k_x,k_y,-k_z\right) = \Omega_{x}\left(k_x,k_y,k_z\right)$ \\
 $m_{y}'$ & $x,-y,z$ & $\Omega_{y}\left(-k_x,k_y,-k_z\right) = -\Omega_{y}\left(k_x,k_y,k_z\right)$ \\
 &  & $\Omega_{z}\left(-k_x,k_y,-k_z\right) = \Omega_{z}\left(k_x,k_y,k_z\right)$ \\
\hline
 &  & $\Omega_{x}\left(k_x,k_y,-k_z\right) = -\Omega_{x}\left(k_x,k_y,k_z\right)$ \\
 $m_{z}$ & $x,y,-z$ & $\Omega_{y}\left(k_x,k_y,-k_z\right) = -\Omega_{y}\left(k_x,k_y,k_z\right)$ \\
  &  & $\Omega_{z}\left(k_x,k_y,-k_z\right) = \Omega_{z}\left(k_x,k_y,k_z\right)$ \\
\hline
  &  & $\Omega_{x}\left(k_y,k_x,-k_z\right) = -\Omega_{y}\left(k_x,k_y,k_z\right)$ \\
 $m_{-xy}'$ & $-y,-x,z$ & $\Omega_{y}\left(k_y,k_x,-k_z\right) = -\Omega_{x}\left(k_x,k_y,k_z\right)$ \\
 &  & $\Omega_{z}\left(k_y,k_x,-k_z\right) = \Omega_{z}\left(k_x,k_y,k_z\right)$ \\
\hline
 &  & $\Omega_{x}\left(-k_y,-k_x,-k_z\right) = \Omega_{y}\left(k_x,k_y,k_z\right)$ \\ 
$m_{x-y}'$ & $y,x,z$ & $\Omega_{y}\left(-k_y,-k_x,-k_z\right) = \Omega_{x}\left(k_x,k_y,k_z\right)$ \\
 &  & $\Omega_{z}\left(-k_y,-k_x,-k_z\right) = \Omega_{z}\left(k_x,k_y,k_z\right)$ \\
\hline
\caption{Transformation of Berry curvature components under the symmetry operations of Magnetic Laue group $4/mm'm'$}
\end{longtable}
\end{tiny}
